%% file: article_4cell_5_11_08.tex
\documentclass[12pt]{article}
\usepackage{amsmath,amsfonts,amscd,amssymb,latexsym,graphicx,epsfig,color,verbatim}
\usepackage[all]{xy}

\newtheorem{definition}{Definition}
\newtheorem{theorem}{Theorem}

\newtheorem{remark}{Remark}
\newenvironment{proof}[1][Proof]{\textbf{#1.} }{\ \rule{0.5em}{0.5em}}

\begin{document}

\title{Heteroclinic Ratchets in a System of Four Coupled Oscillators}

\author{ {\"O}zkan Karabacak and Peter Ashwin\\
Mathematics Research Institute,\\
School of Engineering, Computing and Mathematics,\\
University of Exeter,\\
Exeter EX4 4QF, UK.}

\maketitle

\begin{abstract}
We study an unusual but robust phenomenon that appears in an example system of
four coupled phase oscillators. The coupling is preserved under only one symmetry, 
but there are a number of invariant subspaces and degenerate bifurcations forced by
the coupling structure, and we investigate these. We show that the system can
have a robust attractor that responds to a specific detuning $\Delta$ between
certain pairs of the oscillators by a breaking of phase locking for arbitrary 
$\Delta>0$ but not for $\Delta\leq 0$. As the dynamical mechanism behind this is a 
particular type of heteroclinic network, we call this a `heteroclinic ratchet' 
because of its dynamical resemblance to a mechanical ratchet.
\end{abstract}
\begin{center}

 \end{center}

\section{Introduction}

Coupled oscillators arise as simplified models for coupled limit cycle oscillators in case of weak coupling \cite{strogatz_00}. They have been receiving an increasing interest not only because of their various application areas such as electrochemical oscillators \cite{zhai_kiss_05,kiss_07} and neural systems \cite{rabinovich_review_06} but also because they present analytically tractable models to understand various kinds of dynamical phenomena \cite{kuramoto,hansel_mato_meunier_93,kori_kuramoto_01}. These include complete phase synchronization, partial synchronization due to the existence of stable synchronized clusters and slow switching between unstable clusters. The last phenomenon takes place if there is an attractor composed of unstable cluster states which are connected to each other by heteroclinic connections and thus form a heteroclinic network in state space. 

Heteroclinic networks (or heteroclinic cycles in particular) are used to explain slow switching behaviour of physical systems where a system stays near a dynamically unstable equilibrium or periodic orbit for a long period, then changes its state to another stationary state relatively fast, and repeats this process for another or same stationary state. Despite the fact that heteroclinic networks are not structurally stable, they can be robust if the system considered is constrained by some conditions, such as symmetry \cite{krupa_97,golubitsky_stewart}. This is due to the existence of invariant subspaces on which heteroclinic connections between saddle equilibria can exist robustly. 

This robust behaviour was first observed in examples of rotating convection and explained by the existence of robust heteroclinic cycle in \cite{busse_79} and \cite{guckenheimer_holmes_88}. Heteroclinic networks are used to explain slow switching phenomenon in different areas such as population dynamics \cite{hofbauer_sigmund_98}, electrochemical oscillators \cite{kiss_07,zhai_kiss_05} and neural systems \cite{rabinovich_06,rabinovich_review_06}. They also may have some applications in computational engineering as some recent works \cite{ashwin_borresen_04,ashwin_borresen_05,ashwin_gabor_07} suggest. Especially in complex neural systems, the use of heteroclinic networks are quite promising since this means one can model persistent transient behaviour \cite{rabinovich_review_06}. 

 In case of full permutation symmetry (all-to-all coupling), a system of $N$ coupled oscillators can admit robust heteroclinic networks for $N=4$ or greater \cite{ashwin_burylko_maistrenko_08,ashwin_orosz_borresen_08}. It is important to note that due to the symmetry these heteroclinic networks cannot have arbitrary forms. On the other hand, symmetry is not necessary for robust heteroclinic networks to exist. For example, in \cite{aguiar_ashwin_dias_field_08} it is shown that robust heteroclinic cycles can exist for coupled cell systems with nonsymmetric coupling structure. In this work, we study a coupled phase oscillator system for which robust heteroclinic networks appear in the phase difference space as a result of the coupling structure rather than the symmetry of the coupling. This gives rise to heteroclinic networks with some properties that are not seen for symmetric system. 

We emphasize a new type of heteroclinic network that we call heteroclinic ratchet as it resembles a mechanical ratchet, a device that allows rotary motion on applying a torque in one direction but not in the opposite direction. A heteroclinic ratchet on an $N$-torus contains heteroclinic cycles winding in some directions but no other heteroclinic cycles winding in the opposite directions. We show that this type of heteroclinic network can exist as an attractor in phase space resulting in noise induced desynchronization of certain couples of oscillators in such a way that one of the oscillators has larger frequency than the other for all initial conditions close to synchrony state. This phenomenon is new in coupled phase oscillator systems and cannot take place in all-to-all coupled systems, since the permutation symmetry enforces the system to have desynchronization of a couple, if there is any, in both ways. We will show that the existence of heteroclinic ratchets for a coupled phase oscillator system is mainly related to the coupling structure. Moreover, heteroclinic ratchets have important dynamical consequences such as sensitivity to detuning and noise.

The main model for coupled phase oscillators is the Kuramoto model of $N$ oscillators where each oscillator is coupled
to all the others by a specific $2\pi$-periodic coupling function
\cite{kuramoto}. We consider the same model with a specific connection structure
and using a more general coupling function $g(x)$. Each oscillator has dynamics
given by 
\begin{equation}
 \dot\theta_i=\omega_i+\frac{K}{N}\sum_{j=1}^{N}c_{ij}g(\theta_i-\theta_j).
\label{eqnoscillator}
\end{equation} 
Here $\dot\theta_i\in\mathbb T=[0,2\pi)$ and $w_i$ is the natural frequency of
the oscillator $i$. The connection matrix $\{c_{ij}\}$ represents the coupling
between oscillators. $c_{ij}=1$ if the oscillator $i$ receives an input from the
oscillator $j$ and $c_{ij}=0$ otherwise. The coupling function $g$ is a
$2\pi$-periodic function. For weakly coupled oscillators it is well know that (\ref{eqnoscillator}) will have an $\mathbb T^1$ phase shift symmetry, that is the dynamics of (\ref{eqnoscillator}) are invariant under the phase shift 
$$
(\theta_1,\theta_2,\dots,\theta_N)\mapsto (\theta_1+\epsilon,\theta_2+\epsilon,\dots,\theta_N+\epsilon)
$$
for any $\epsilon\in \mathbb T$. We will initially consider identical oscillators, 
that is, 
\begin{equation}
 \label{eq_tuning}
\omega_i=\omega\ ,\ \ i=1,\dots,N
\end{equation}
before discussing at a latter stage the effect of detuning where the
oscillators can have different natural frequencies. 
Because the coupling function $g$ is $2\pi$-periodic it is natural to consider a Fourier series expansion
\begin{equation}
 g(x)=\sum_{k=1}^{\infty} r_k\sin (kx+\alpha_k)
\label{eqncoupling}
\end{equation}
where $r_k$ must converge to zero fast enough and $\alpha_k$'s are arbitrary.
Several truncated cases of the general case (\ref{eqncoupling}) have been considered in the literature:
\begin{itemize}
\item
Setting $r_k=0$ for $k=2,3,\dots$ and $\alpha_1=0$ gives the Kuramoto model, which exhibits frequency synchronization and clustering phenomena \cite{kuramoto}. 
\item
Setting $r_k=0$ for $k=2,3,\dots$ but leaving arbitrary $\alpha_1$ gives the Kuramoto-Sakaguchi model \cite{sakaguchi_kuramoto_86} with essentially the same dynamics as the Kuramoto model. 
\item
Setting $r_k=0$ for $k=3,4,\dots$ and $\alpha_2=0$ gives the model of Hansel et al. \cite{hansel_mato_meunier_93}. They showed that one can observe new phenomena not present in the above cases. For example, taking 
$$
r_1=-1,\ r_2=0.25,\ a_1=1.25
$$
and setting all other parameters to zero, they show that one can observe slow 
switching phenomenon as a result of the presence of an asymptotically stable 
robust heteroclinic cycle connecting a pair of saddles.
\end{itemize}

We investigate a particular four-coupled cell system that admits a robust heteroclinic ratchet as an attractor only in presence of a third harmonic in the coupling 
function, i.e. we will require $r_3\neq 0$. 
%\marginpar{Not sure if this makes the exposition more simple?}
 %For simplicity, we first take $r_3=\alpha_2=0$ and consider only two parameters $\alpha=\alpha_1$ and $r=r_2$ while showing the existence of robust heteroclinic networks and then consider negative values of $\alpha_3$ in order to obtain an attracting heteroclinic network. 
Note that, without loss of generality, we also set $K=N$ and $r_1=-1$ by a scaling of time.

The coupling structure considered in this work (see Figure \ref{fig4cell}) arises as an inflation of the all-to-all coupled 3-cell network \cite{aguiar_ashwin_dias_field_08}. As the network admits an $S_3$-symmetric quotient network there may exist symmetry broken branches of solutions for the coupled systems associated to this network \cite{aguiar_dias_golubitsky_leite_07}. This is a direct result of the Equivariant Branching Lemma \cite{golubitsky_stewart}. We will show that for the coupled oscillator system we consider, such a synchrony breaking bifurcation includes two extra pitchfork branches as a result of the $\mathbb T^1$-phase shift symmetry. These correspond to the saddle cluster states which may form heteroclinic ratchets for some parameter region. 

This work consist of three parts. In Section \ref{sec_systems}, we will analyze the dynamics of the coupled cell system of four phase oscillators and find the invariant subspaces where heteroclinic networks
can exist. Theorem 1 characterizes a syncrony breaking bifurcation in such systems. In Section \ref{sec_ratchets}, we consider a particular coupling function and explain the emergence of heteroclinic ratchet connecting two pitchfork branches given in Theorem 1. Finally in Section~\ref{sec_consequences}, we discuss dynamical consequences of the heteroclinic ratchet, considering the influence of noise and detuning of natural frequencies.
%Considering a synchrony breaking steady state bifurcation of the
%origin we explain the emergence of heteroclinic networks of different types.
%These different behaviour of oscillator system can be considered as different
%synchronization phenomena. In Section 3, examples of such different
%synchronization types are given considering also the presence of small noise.
%Finally, we will show that for an open set of parameters oscillators are
%extremely sensitive to detuning due to the existence of a one-way heteroclinic
%network.
\section{An example of four coupled oscillators}
\label{sec_systems}
\begin{figure}
\begin{center}
\input{4cell.pstex_t} 
\end{center}
\caption{
\sc A 4-cell network: this gives coupled systems of the form (\ref{eq_4cellrighthandsidefunction}). Observe that the network has a single symmetry given by the permutation $\left(12\right)\left(34\right)$.}
\label{fig4cell}
\end{figure}
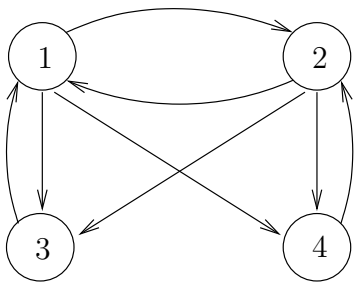

In this section, we consider a specific case of four
identical oscillators coupled by a connection structure shown
in Figure \ref{fig4cell}. More specifically, the system we consider is
\begin{equation}
\begin{split}
\dot\theta_1&=\omega_1+f(\theta_1;\theta_2,\theta_3)\\
\dot\theta_2&=\omega_2+f(\theta_2;\theta_1,\theta_4)\\
\dot\theta_3&=\omega_3+f(\theta_3;\theta_1,\theta_2)\\
\dot\theta_4&=\omega_4+f(\theta_4;\theta_1,\theta_2).
\end{split}
 \label{eq4cell4dimension}
\end{equation}
We first consider identical oscillators, that is 
\begin{equation}
\label{eq_identity}
	\omega=\omega_1=\cdots=\omega_4.
\end{equation}
Oscillators with different natural frequencies will be considered in Section~\ref{sec_consequences}.
We assume that the inputs to each cell are indistinguishable, i.e.
\begin{equation}
\label{eq_inputsymmetryoff}
f(x;y,z)=f(x;z,y)\ \ \text{for all } x,y,z\in{\mathbb T}.
\end{equation}
We will also assume the presence of the phase shift symmetry
\begin{equation}
f(x+\epsilon;y+\epsilon,z+\epsilon)=f(x;y,z)\ \ \text{for all } x,y,z,\epsilon\in
\mathbb T.
 \label{eq_phaseshiftsymmetryoff}
\end{equation}
This $\mathbb T^1$-symmetry arises for example in weakly coupled limit cycle oscillators via averaging \cite{ashwin_swift_92}.
Note that, for the present section, the form of coupling we assume will be more general than (\ref{eqnoscillator}).

%The system has far fewer symmetries than for example present in an all-to-all coupled system of four oscillators. 
In the following we discuss the invariant subspaces of (\ref{eq4cell4dimension}) and give a result about the solution branches on the invariant subspaces that emanate at bifurcation from a fully synchronized solution.

\subsection{Invariant subspaces}

The network in Figure \ref{fig4cell} has a symmetry that we characterize as follows. Let $\Gamma$ be an $S_2$-action on $\mathbb T^4$ generated by
$$
\sigma\colon(\theta_1,\theta_2,\theta_3,\theta_4)\rightarrow
(\theta_2,\theta_1,\theta_4,\theta_3).
$$ 
The symmetry of the network implies that the system (\ref{eq4cell4dimension}) is $\Gamma$-equivariant and the
fixed point subspace of $\Gamma$, that is, $$\text{Fix}(\Gamma)=\lbrace x\in
\mathbb T^4\mid \sigma x=x \text{ for all } \sigma\in\Gamma\rbrace$$ is invariant under
the dynamics of (\ref{eq4cell4dimension}). Note that, because $\Gamma$ is defined on a torus, $\text{Fix}(\Gamma)$ consists of two disjoint subsets each of
which is invariant: $V^{s3}_2=\lbrace\theta\in\mathbb
T^4\mid\theta_1=\theta_2,\theta_3=\theta_4\rbrace$ and $\bar
V^{s3}_2=\lbrace\theta\in
T^4\mid\theta_1=\theta_2+\pi,\theta_3=\theta_4+\pi\rbrace$. On the other hand,
there are many other invariant subspaces which do not appear because of the symmetries of the network but because of the groupoid structure of the input sets
of cells (see \cite{golubitsky_stewart_06} for groupoid formalism). 

The invariant subspaces can be obtained using the balanced coloring method. A coloring of cells, that is, a partition of the set of all cells into a number of groups or colors is called \emph{balanced} if each pair of cells with same color receive same number of inputs from the cells with any given color. Each balanced coloring gives rise to an invariant subspace obtained by equalizing the state of cells with same color.
Moreover, each balanced coloring corresponds to a quotient network which gives the dynamics reduced to the corresponding invariant subspace. 

\begin{table}
\centering
	\begin{tabular}{|c|l|}
		\hline
		Dimensions & Invariant Subspaces\\
		\hline
		4 & $V_4=\mathbb T^4$\\
		3 & $V_3^s=\{\mathbf{\theta}\in \mathbb T^4\mid \theta_3=\theta_4\}$\\
		3 & $V_3^{1}=\{\mathbf{\theta}\in \mathbb T^4\mid \theta_2=\theta_4\}$\\
		3 & $V_3^{2}=\{\mathbf{\theta}\in \mathbb T^4\mid \theta_1=\theta_3\}$\\
		2 & $V_2=\{\mathbf{\theta}\in \mathbb T^4\mid \theta_1=\theta_3,
\theta_2=\theta_4\}$\\
		2 & $V_2^{s1}=\{\mathbf{\theta}\in \mathbb T^4\mid \theta_2=\theta_3=\theta_4\}$\\
		2 & $V_2^{s2}=\{\mathbf{\theta}\in \mathbb T^4\mid \theta_1=\theta_3=\theta_4\}$\\
		2 & $V_2^{s3}=\{\mathbf{\theta}\in \mathbb T^4\mid \theta_1=\theta_2,\theta_3=\theta_4\}$\\
		1 & $V_1=\{\mathbf{\theta}\in \mathbb T^4\mid
\theta_1=\theta_2=\theta_3=\theta_4\}$\\
		\hline
	\end{tabular}
\caption{
\label{table_invariantsubspaces}
\sc Invariant subspaces forced by the coupling structure in Figure \ref{fig4cell} for the system (\ref{eq4cell4dimension})}
\end{table}

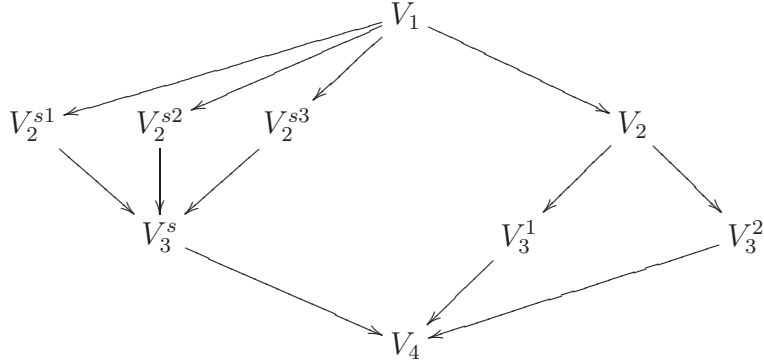
\begin{figure}[ht]
\begin{center}
$$
\xymatrix{ & &  &V_1\ar[dll]\ar[dl]\ar[dlll]\ar[drr] & & & \\
V_2^{s1}\ar[dr] & V_2^{s2}\ar[d] & V_2^{s3}\ar[dl] & & &  V_2\ar[dl]\ar[dr] & \\
& V_3^s\ar[drr] & & & V_3^1\ar[dl] & & V_3^2\ar[dlll]\\ 
 & & &V_4& & &
}
$$
\end{center} 
\caption{
\sc Containment of the invariant subspaces given in Table \ref{table_invariantsubspaces}. $V_x\rightarrow V_y$ means $V_x\subset V_y$. The subscripts indicates the dimensions of the invariant subspaces and the superscript $s$ labels the fixed point subspaces related to the $S_3$-symmetry of the quotient network for $\theta_3=\theta_4$.}
\label{figordering}
\end{figure}

For the system (\ref{eq4cell4dimension}) the invariant subspaces obtained by the balanced coloring method are listed in Table~\ref{table_invariantsubspaces}. The subscripts indicate dimensions of subspaces and there exist a partial ordering for the set of these subspaces given by containment, that is, $$V_x\prec V_y \Leftrightarrow V_x\subset V_y.$$ This ordering of invariant subspaces is illustrated in Figure \ref{figordering}.

Consider the balanced coloring $\{3,4\}$, where only third and forth cells have same color. The corresponding invariant subspace is $V_3^s$ and the quotient network is the $S_3$-symmetric all-to-all coupled 3-cell network (see Table \ref{table_quotients}). Necessarily all the fixed point subspaces of this 3-cell quotient lift to some invariant subspaces of the 4-cell system and these are labelled by the superscript $s$.
Note that $V_2^{s3}$ is the only one of these that is contained in
$\text{Fix}(\Gamma)$, but there are some pairs of subspaces for which one
subspace is related to the other by the symmetry of the system, namely $\sigma(V_2^{s1})=V_2^{s2}$ and $\sigma(V_3^1)=V_3^2$. As a result, the quotient networks corresponding the subspaces $V_3^1$ and $V_3^2$ are also symmetrically related (see Table \ref{table_quotients}).

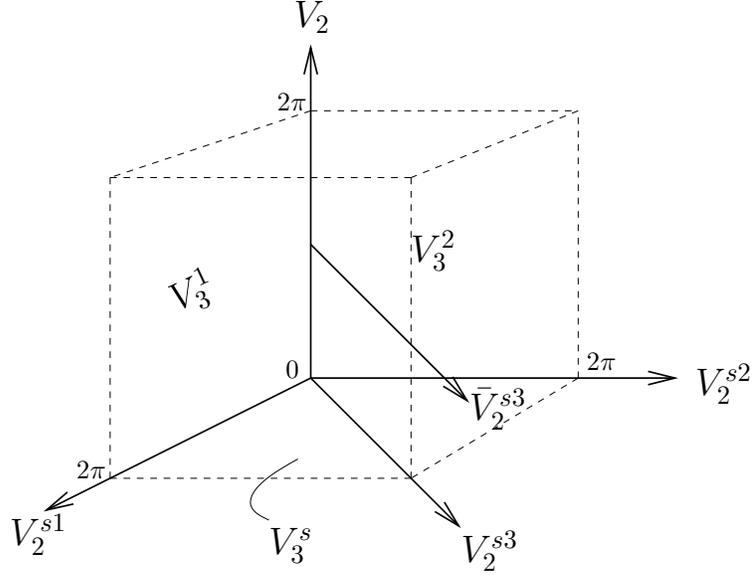
\begin{figure}[ht]
\begin{center}
\input{invariantsubsets.pstex_t}
\end{center} 
\caption{
\sc Invariant subspaces given in Table 1 projected onto $\mathbb T^3$ (represented by a unit cell in $\mathbb R^3$). Subscripts indicates
the subspace dimensions on $\mathbb T^4$)}
\label{figinvariantsubspaces}
\end{figure}

Exploiting the phase shift symmetry (\ref{eq_phaseshiftsymmetryoff}), the 4-dimensional system (\ref{eq4cell4dimension}) and (\ref{eq_identity}) can be reduced to a 3-dimensional one by defining new variables 
$$
(\phi_1,\phi_2,\phi_3):=(\theta_1-\theta_3,\theta_2-\theta_4,\theta_3-\theta_4)
$$
so that
\begin{eqnarray}
\dot{\phi_1}&=& f(\phi_1;\phi_2-\phi_3,0)
-f(0;\phi_1,\phi_2-\phi_3)\nonumber\\
\dot{\phi_2}&=& f(\phi_2;\phi_1+\phi_3,0) -f(0;\phi_1+\phi_3,\phi_2)\label{eqndphidt}\\
\dot{\phi_3}&=& f(\phi_3;\phi_1+\phi_3,\phi_2)
-f(0;\phi_1+\phi_3,\phi_2).\nonumber
\end{eqnarray}
The symmetry of the system (\ref{eq4cell4dimension}) has 
implications for this system. Let $\tilde \Gamma$ be an $S_2$-action
on $\mathbb T^3$ generated by $\rho:(\phi_1,\phi_2,\phi_3)\rightarrow
(\phi_2,\phi_1,-\phi_3)$. Then the system (\ref{eqndphidt}) is $\tilde \Gamma$ equivariant. In this case the fixed point subspaces are the lines $\{\phi\in\mathbb
T^3\mid\phi_1=\phi_2,\ \phi_3=0\}$ and $\{\phi\in\mathbb T^3\mid\phi_1=\phi_2,\
\phi_3=\pi\}$. Other invariant subspaces can be obtained projecting the
previously found invariant subspaces onto $\mathbb T^3$. These are illustrated
in Figure~\ref{figinvariantsubspaces}.

\begin{table}
\centering
\begin{tabular}{|c|c|c|}
\hline
Balanced & Invariant & Quotient\\
Colourings & Subspaces & Networks\\
\hline
& & \\
\raisebox{8.5ex}{$\{3,4\}$} & \raisebox{8ex}{$V_3^s$} &\input{3cellV3a.pstex_t}\\
\raisebox{8.5ex}{$\{2,4\}$} & \raisebox{8ex}{$V_3^1$}
&\input{3cellV3b2.pstex_t}\\
\raisebox{8.5ex}{$\{1,3\}$} & \raisebox{8ex}{$V_3^2$}
&\input{3cellV3b1.pstex_t}\\
\hline
\end{tabular}
\caption{
\sc Quotient Networks for the Three Dimensional Invariant Subspaces $V_3^s$, $V_3^1$, and $V_3^2$ of the 4-cell system (\ref{eq4cell4dimension})}
\label{table_quotients}
\end{table} 

\subsection{Synchrony breaking bifurcations}

In \cite{aguiar_dias_golubitsky_leite_07}, it is shown that any coupled
cell system that has a connection structure as in Figure \ref{fig4cell} admits
an $S_3$-transcritical bifurcation on $V_3^s$ at the origin. More concretely,
there exist three transcritical branches of unstable solutions on $V_2^{s1}$,
$V_2^{s2}$, and $V_2^{s3}$ simultaneously emanating from the origin if
$f_x(0)-f_y(0)=0$ and some transversality inequalities are satisfied. However,
for the coupled phase oscillators of type (\ref{eq4cell4dimension}), apart from the
connection structure, dynamical properties affect the bifurcation scheme. Now we
will show in Theorem \ref{theorem} how the $\mathbb T^1$-symmetry of $f$ gives rise to a pitchfork bifurcation on
$V_2$ that takes place simultaneously with the transcritical bifurcations
mentioned above. The occurrence of simultaneous branches on invariant lines is
not only a consequence of the Equivariant Branching Lemma
\cite{golubitsky_stewart} but also a result of the connection structure and the
property of the individual dynamics, that is the $\mathbb T^1$-symmetry of $f$.

\begin{table}
\centering
\begin{tabular}{|c|c|}
\hline
Adjacency matrix & eigenvalues and eigenvectors\\
\hline
$A=\left( \begin{array}{cccc}
0 & 1 & 1 & 0 \\
1 & 0 & 0 & 1 \\
1 & 1 & 0 & 0 \\
1 & 1 & 0 & 0 \end{array} \right)$ &
\begin{tabular}{c c}
$\mu_1=-1$, & $\nu_1=(1,-1,0,0)^\text T$\\
$\mu_2=-1$, & $\nu_2=(0,-1,1,1)^\text T$\\
$\mu_3=0$, & $\nu_3=(1,-1,1,-1)^\text T$\\
$\mu_4=2$, & $\nu_4=(1,1,1,1)^\text T$\\
\end{tabular}\\
\hline
\end{tabular}
\caption{
\sc Adjacency matrix of the network in Figure \ref{fig4cell} with eigenvalues and eigenvectors}
\label{table_adjacency}
\end{table} 

\begin{theorem}\label{theorem}
Assume that $\alpha$ is a parameter of the system (\ref{eqndphidt}). If $f$ satisfies 
$f_x(0,\alpha^*)=0$, $f_{x\alpha}(0,\alpha^*)\neq 0$, and $f_{xxx}(0,\alpha^*)\neq 0$ then there 
exist a pitchfork bifurcation of the origin of ({\protect\ref{eqndphidt}}) on $V_2$ at 
$\alpha=\alpha^*$ appearing simultaneously with the transcritical bifurcations on 
$V_2^{s1}$, $V_2^{s2}$ and $V_2^{s3}$.
\end{theorem}

\begin{remark}
A direct consequence of the Theorem \ref{theorem} is that a generic bifurcation of the fully
synchronized periodic solution $(x,x,x,x)$ of (\ref{eq4cell4dimension}) will give rise to three branches of periodic solutions of the form
\begin{equation*}\begin{array}{l}
(x,y,x,x)\\
(y,x,x,x)\\
(x,x,y,y)\\
\end{array}
\end{equation*}
and two other branches of the form $
(x,y,x,y)$,
where the first three appear by transcritical bifurcations and the final two via a pitchfork bifurcation.
\end{remark}

\begin{proof}
Consider the adjacency matrix $A$ of the network (see Table \ref{table_adjacency}). The eigenvalues of $A$ and partial derivatives of $f_x$ and $f_y$ ($f_z=f_y$) at the
origin determine the stability of the origin (see Proposition 2 in
\cite{aguiar_dias_golubitsky_leite_07}). The eigenvalues of (\ref{eqndphidt}) at the origin are
\begin{equation}
 \lambda_i=f_x(0,\alpha)+\mu_if_y(0,\alpha)
\label{eqneigenvalues}
\end{equation}
where $\mu_i$ is an eigenvalue of $A$ and $i=1,2,3$. The eigenvectors of (\ref{eqndphidt}) are
the same as the $A$'s. It is important to note that the $\mathbb T^1$ phase shift symmetry of (\ref{eq4cell4dimension}) induce a relation between partial derivatives:
\begin{equation}
 f_x+f_y+f_z\equiv 0.
\label{eqnpartialderivatives}
\end{equation}
This can be obtained taking the derivative of (\ref{eq_phaseshiftsymmetryoff}) with respect to $\epsilon$, and (\ref{eq_inputsymmetryoff}) implies $f_y\equiv f_z$ Thus, there exist a linear relationship between partial derivatives
\begin{equation}
 f_x\equiv-2f_y.
\label{eq_fxfy}
\end{equation}
Derivatives of (\ref{eqnpartialderivatives}) with respect to $x$ and $y$ give
\begin{eqnarray}
 f_{xx}\equiv-2f_{yx}\equiv4f_{yy}
\label{eqnpartialderivativesextras1}\\
 f_{xxx}\equiv-2f_{yxx}.
\label{eqnpartialderivativesextras2}
\end{eqnarray}
The Eq. \ref{eqneigenvalues} and \ref{eq_fxfy} imply that the eigenvalues
$\lambda_i$ become zero simultaneously for all invariant lines passing through
the origin when $f_x(0,\alpha)=0$. To see that there exist a pitchfork branch on
$V_2$ we consider the solutions of type $(x,x+u/2,x,x+u/2)$.
Substituting this into (\ref{eqndphidt}) and using the Eq. \ref{eq_phaseshiftsymmetryoff}, one gets 
$\dot u=F(u):=f(0,0,- u,\alpha)-f(0, u,0,\alpha)$. Thus the assumptions $f_{x\lambda}(0,\alpha^*)\neq 0$, $f_{xxx}(0,\alpha^*)\neq 0$ and the Eq.'s \ref{eq_fxfy}, \ref{eqnpartialderivativesextras1} and \ref{eqnpartialderivativesextras2} imply the pitchfork bifurcation conditions $(\partial^2F/\partial u^2)(0,\alpha^*)=0$, $(\partial^2F/\partial\lambda\partial u)(0,\alpha^*)\neq 0$ and $(\partial^3F/\partial u^3)(0,\alpha^*)\neq 0$. Since these also imply the assumptions of Theorem 1 in \cite{aguiar_dias_golubitsky_leite_07}, there exist simultaneous transcritical bifurcations on $V^{s1}_2$, $V^{s2}_2$ and $V^{s3}_2$.
\end{proof}
\section{Robust heteroclinic ratchets for the system of four coupled oscillators}
\label{sec_ratchets}
In the previous section, it is shown that the connection structure of the system (\ref{eq4cell4dimension}) induce the existence of invariant subspaces. These subspaces persist under the perturbations that preserve the connection structure. For this reason, as in symmetric systems, one can find robust heteroclinic networks laying on the invariant subspaces of the system (\ref{eq4cell4dimension}). By ``robust" we mean the persistence under the small perturbations that preserve the coupling structure. We will see that for the phase difference system (\ref{eqndphidt}) some unusual heteroclinic networks exist, which are not seen for symmetric systems. We distinguish one type of these heteroclinic networks, which we call as heteroclinic ratchet because it includes connections that wind around the torus in one direction only.
\begin{definition}
 For a system on $\mathbb T^N$, an invariant set is a \emph{heteroclinic ratchet} if it includes a heteroclinic cycle with nontrivial winding in one direction but no heteroclinic cycles winding in the opposite direction. More precisely, we say a heteroclinic cycle $C$ parametrized by $x(s), s\in[0,1)$ has \emph{nontrivial winding in some direction} if  there is an angular variable $P\colon\mathbb T^N\rightarrow\mathbb R$ with $P(\mathbb T^N)=[0,2\pi)$ and $P(x(0))=0$ such that $\lim_{s\rightarrow 1}P(x(s))-\lim_{s\rightarrow 0}P(x(s))=2\pi$. A heteroclinic cycle winding in the opposite direction would similarly have $\lim_{s\rightarrow 1}P(x(s))-\lim_{s\rightarrow 0}P(x(s))=-2\pi$.
\end{definition}
\begin{remark}
 If a system has a heteroclinic ratchet, then there exist pseudo-orbits winding in one direction on the torus.
\end{remark}

In this section, we will first explain how a heteroclinic ratchet emerges for the system (\ref{eqndphidt}) after a synchrony breaking bifurcation. Then, we will discuss the stability of the heteroclinic ratchet and exhibit a coupling function $g$ for which the heteroclinic ratchet is an attractor. Finally, a heteroclinic network with different structure is shown to exist for other parameter values.
\subsection{Heteroclinic ratchets for the four coupled oscillators}
We consider the particular case of (\ref{eq4cell4dimension}):
\begin{equation}
f(x;y,z)=g(x-y)+g(x-z).
 \label{eq_4cellrighthandsidefunction}
\end{equation}
% We again assume (\ref{eq_identity}). 
% The case of nonidentical natural frequencies will be considered in the next section.

Using (\ref{eq_4cellrighthandsidefunction}), we can write the phase difference system with identical natural frequencies given in Eq. \ref{eqndphidt} in the form 
\begin{eqnarray}
\label{eqn3cellspecific}
\dot{\phi_1}&=&g(\phi_1+\phi_3-\phi_2)+g(\phi_1)-g(-\phi_1)-g(\phi_3-\phi_2)\nonumber\\
\dot{\phi_2}&=&g(\phi_2-\phi_3-\phi_1)+g(\phi_2)-g(-\phi_3-\phi_1)-g(-\phi_2)\\
\dot{\phi_3}&=&g(-\phi_1)+g(\phi_3-\phi_2)-g(-\phi_3-\phi_1)-g(-\phi_2).\nonumber
\end{eqnarray}
Here the new variables are the phase differences $\phi_1:=\theta_1-\theta_3$,
$\phi_2:=\theta_2-\theta_4$, and $\phi_3:=\theta_3-\theta_4$. We consider the coupling function $g$ up to three harmonics:
\begin{equation}
\label{couplingfunction}
 g(x)=-\sin (x+\alpha_1)+r_2\sin (2x)+r_3\sin (3x).
\end{equation}
For this coupling function, there may exist different types of robust heteroclinic networks for different parameter values. We first demonstrate a heteroclinic ratchet that exists for an open set of parameters.

\begin{figure}
\begin{center}
\input{heteroclinic_ratchet_all_in_one.pstex_t}
\end{center} 
\caption{
\sc Heteroclinic Ratchet for the system (\ref{eqn3cellspecific}) with the parameter set (\ref{eq_parameterset1}). Sources, saddles and sinks are
indicated by small disks filled with white, grey or black color. (a) Phase portrait on $V_3^1$. The horizontal axis is $V_2^{s1}$ ($\phi_1$-axis)
and the vertical axis is $V_2$ ($\phi_3$-axis). (b) The heteroclinic ratchet on the boundary of $\mathbb T^3$. (c) The heteroclinic ratchet seen on the lift of $\mathbb T^3$ to $\mathbb R^3$.}
\label{fig_heteroclinic_ratchet}
\end{figure}
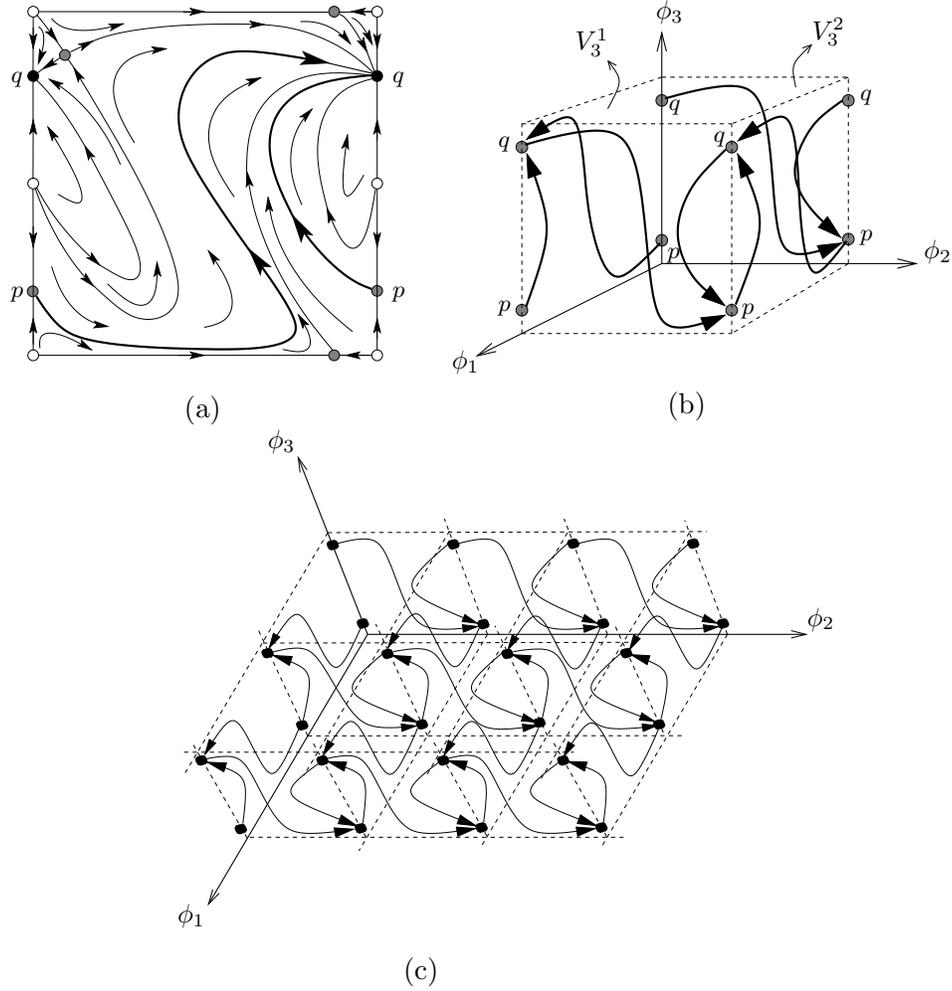

Heteroclinic networks are usually exceptional phenomena, but they can be robust if the associated heteroclinic connections lie within invariant subspaces \cite{krupa_97}. For (\ref{eqn3cellspecific}) and (\ref{couplingfunction}) there are invariant subspaces that are found in the previous section for a more general system (\ref{eqndphidt}) (see Figure \ref{figinvariantsubspaces}). For the parameter set
\begin{equation}
\label{eq_parameterset1}
(\alpha_1,r_2,r_3)=(1.4,0.3,-0.1)	
\end{equation}
we identify robust heteroclinic connections from the equilibrium $p$ to $q=\sigma (p)=2\pi -p$ on one of these invariant subspaces, namely $V_3^1$, using the simulation tool XPPAUT \cite{ermentrout} (see Figure
\ref{fig_heteroclinic_ratchet}a). Recall that the subspaces $V_3^1$ and $V_3^2$ are mapped to each other by the symmetry $\sigma$. Thus, the presence of a connection from $p$ to $q$ on $V_3^1$ implies the presence of another connection on $V_3^2$ that connects $q$ to $p$. Therefore, a heteroclinic network exists on $\mathbb T^3$ for the parameter set (\ref{eq_parameterset1}) (see Figure \ref{fig_heteroclinic_ratchet}b). Note that this is a heteroclinic ratchet since it includes phase slips in the directions $+\phi_1$ and $+\phi_2$.

The equilibria $p$ and $q$ lie on $V_2=V_3^1\cup V_3^2$ and they bifurcate from the origin via a pitchfork bifurcation simultaneously with other transcritical branches of solutions on $V_2^{s1}$, $V_2^{s2}$ and $V_2^{s3}$. This synchrony breaking bifurcation is discussed in Theorem \ref{theorem}. Although we cannot rule out the possibility of the presence of more complex behaviours near this bifurcation, we numerically find the heteroclinic ratchet for the parameter values close to the bifurcation point. This suggests that the bifurcation given in Theorem \ref{theorem} may be associated with a global bifurcation to a heteroclinic ratchet.

The heteroclinic ratchet found for the parameter set (\ref{eq_parameterset1}) is an asymptotically stable attractor.
The asymptotic stability of robust heteroclinic networks can be observed
by considering the transversal, contracting and expanding eigenvalues at the
equlibria $p$, that is, $\lambda_t(p)$, $\lambda_c(p)$, and $\lambda_e(p)$
respectively \cite{krupa_melbourne_95}. Since in our example there is no
transversal direction, and $p$ and $q$ are symmetrically related, a heteroclinic network connecting the equilibria $p$ and $q$ is asymptotically stable if
$|\lambda_e(p)/\lambda_c(p)|<1$ and unstable if
$|\lambda_e(p)/\lambda_c(p)|>1$. For the parameter set (\ref{eq_parameterset1}), the equlibrium is at $p=1.4432$. Then, $\lambda_e(p)$ and $\lambda_c(p)$ are $0.74$ and $-1.2$ by linearizing (\ref{eqn3cellspecific}) at $p$ (see Eq. \ref{eq_lambdas}). This implies asymptotic stability of the heteroclinic ratchet. 

Note that, since the condition for the asymptotic stability is open and the heteroclinic connections are robust, one can find an open set in parameter space $\{(\alpha_1,r_2,r_3)\mid 0\leq r_2,r_3, 0\leq\alpha_1<2\pi\}$, for which the system (\ref{eqn3cellspecific}) admits an asymptotically stable robust heteroclinic ratchet. On the other hand, for the system (\ref{eqn3cellspecific}), the robust heteroclinic ratchet connecting a pair of saddles $p$ and $q$ on $V_3^s$ cannot be asymptotically stable if $r_3=0$ (see Appendix). Therefore, the heteroclinic ratchets for the system (\ref{eqn3cellspecific}) cannot be asymptotically stable unless the third or higher harmonics of the coupling function $g$ are taken into account.
\subsection{Bifurcation between heteroclinic networks}
\begin{figure}[ht]
\begin{center}
\input{bifurcating_ratchet_all_in_one.pstex_t}
\end{center} 
\caption{
\sc Phase Portraits on $V_3^1$ for $r=0.3$ and (a) $\alpha=1.2$, (b)
$\alpha\cong1.315$, and (c) $\alpha=1.4$ demonstrating a bifurcation from a heteroclinic cycle to a heteroclinic ratchet shown in (d) and (e), respectively. 
\footnotesize{(For each graph the horizontal axis is $V_2^{s1}$ ($\phi_1$-axis)
and the vertical axis is $V_2$ ($\phi_3$-axis). Sources, saddles and sinks are
indicated by small disks filled with white, grey or black color. The parts of
the heteroclinic networks are shown by thick lines.)}}
\label{fig_bifurcatingratchet}
\end{figure}
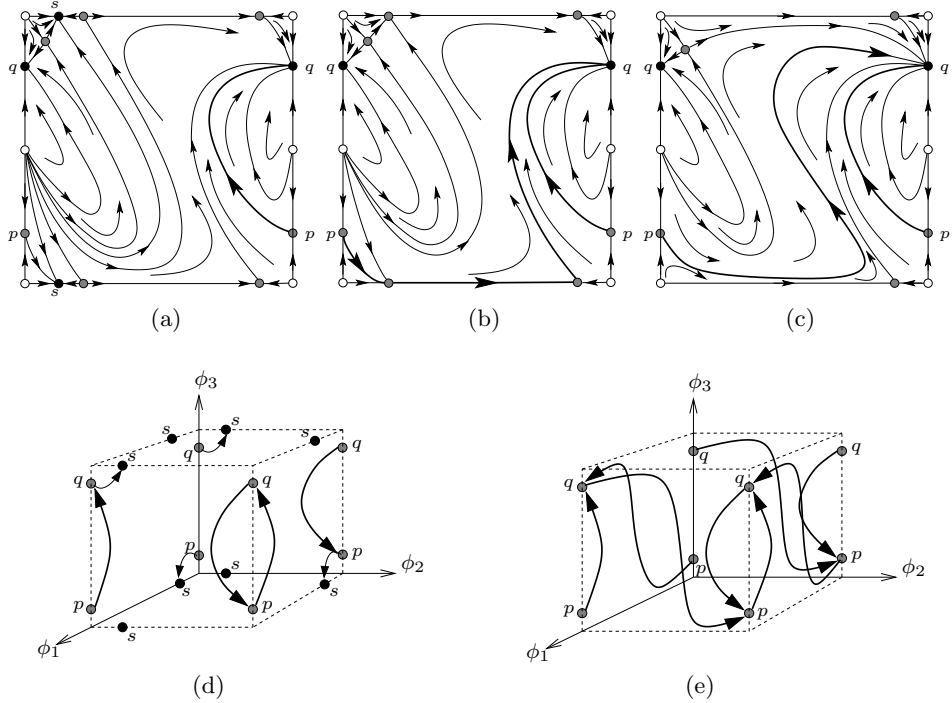

Although the subspace $V_3^s$ does not include any part of the heteroclinic
networks, the dynamics restricted on this subspace, that is, the dynamics of the
network $N_1$ (see Table \ref{table_quotients}) gives rise to a bifurcation from a heteroclinic cycle to a heteroclinic ratchet as seen in Figure \ref{fig_bifurcatingratchet}.
The detailed bifurcation analysis of the 3-cell all-to-all coupled oscillators with a coupling function having the first two harmonics is
given in \cite{ashwin_burylko_maistrenko_08}. There, it is stated that
apart from the transcritical bifurcation of the origin there exists a saddle node
bifurcation on invariant lines. In the reverse direction at this saddle-node
bifurcation, for system (\ref{eqn3cellspecific}) a heteroclinic cycle bifurcates to a heteroclinic ratchet as seen in Figure \ref{fig_bifurcatingratchet}b. 
%\begin{figure}
%\begin{center}
%\input{schematic_hetcyc_most_3D.pstex_t}
%\end{center} 
%\caption{
%\sc Schematic illustration of the robust heteroclinic networks on $\mathbb
%T^3$ for $\alpha=1.2$ (a) and $\alpha=1.3$ (b). $r$ is fixed to $0.2$. Thin
%arrows in (a) show the unstable manifolds of $p$ and $q$ outside the network.}
%\label{schematicont}
%\end{figure}
This bifurcation should also exist for nonzero $r_3$ values and thus one can obtain the heteroclinic cycle in Figure \ref{fig_bifurcatingratchet}d as attracting. In fact, for the parameter set 
\begin{equation}
\label{eq_parameterset2}
(\alpha_1,r_2,r_3)=(1.2,0.3,-0.05),
\end{equation}
the same cycle exists and it turns out to be stable since $\lambda_e(p)=0.68$ and $\lambda_c(p)=-0.70$. Thus the heteroclinic cycle for (\ref{eq_parameterset2}) attracts nearby points with $\phi_1$ and $\phi_2$ less than $2\pi$ and repells other nearby points with $\phi_1$ or $\phi_2$ greater than $2\pi$ to the sink $s$. That is, this heteroclinic cycle has a basin with positive measure, so it is a Milnor attractor, though not stable. This type of a heteroclinic cycle is also unusual for symmetric systems.
\section{Discussion and dynamical consequences of the heteroclinic ratchet}
\label{sec_consequences}

This paper has so far demonstrated that the system of four coupled oscillators in
Figure~\ref{fig4cell} with identical natural frequencies $\omega_i$ can support a robust heteroclinic attractor analogous to a mechanical ratchet. In this section we consider the response of such an attractor to imperfections in the system; in particular we consider
the effect of setting the detunings
$$
\Delta_{ij}=\omega_i-\omega_j
$$
nonzero, and the effect of adding noise to the system. The frequency locking response to detuning and/or noise gives signatures of the presence of the ratchet.

For typical trajectories in terms of the original phases $\theta_i(t)\in \mathbb{R}$ one can define the average frequency of the $i$th oscillator $\Omega_i=\lim_{t\rightarrow \infty}\frac{\theta_i(t)}{t}$ and the frequency difference
$$
\Omega_{ij}=\lim_{t\rightarrow \infty}\frac{\theta_i(t)-\theta_j(t)}{t}.
$$

\begin{definition}
We say the $i$th and $j$th oscillators are {\em frequency synchronized} on an attractor of the system if all trajectories  approaching the attractor satisfy $\Omega_{ij}=0$.
\end{definition}

Note that a stronger notion of synchrony is phase synchronization; we say the
$i$th and $j$th oscillators are {\em phase synchronized} if all trajectories approaching the attractor have $\theta_i(t)-\theta_j(t)$ bounded in $t$. Phase synchronization is a sufficient condition for frequency synchronization, but the converse is not always true as we see below.

\subsection{Response of the system to detuning}

\begin{figure}
\begin{center}
\includegraphics[width=0.60\textwidth]{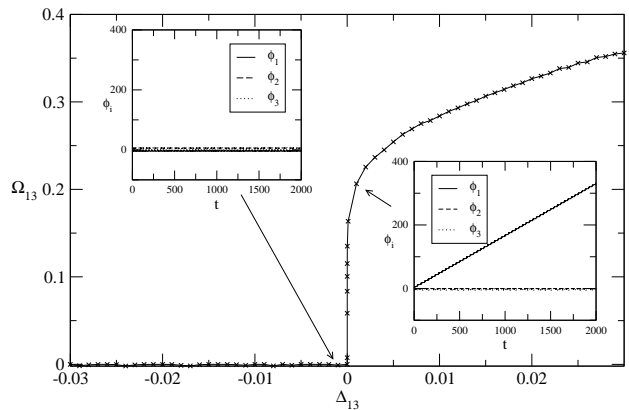}
\end{center} 
\caption{\sc The main graph shows the frequency difference $\Omega_{13}$ for ({\protect
\ref{eq4cell4dimension}}) with parameters ({\protect \ref{eq_parameterset1}}) as a function of detuning $\Delta_{13}$ between the first and third oscillator for $\Delta_{24}=\Delta_{34}=0$. Note that oscillators remain frequency
synchronized for $\Delta_{13}\leq 0$ but quickly break synchrony for $\Delta_{13}>0$;
this is evidence of the attractor being a heteroclinic ratchet. The insets show time evolution
of the phase differences $\phi_i$ for a positive and a negative value of $\Delta_{13}$; observe that oscillators $1$ and $3$ are phase and frequency synchronized for $\Delta_{13}<0$ but neither
phase nor frequency synchronized for $\Delta_{13}>0$.}
\label{fig_detuning}
\end{figure} 

Note that in the case of identical natural frequencies, the oscillators of the original system are frequency synchronized for all trajectories; this follows because trajectories of the reduced phase difference system are trapped inside a bounded invariant region, and so they are phase synchronized. As soon as $\Delta_{ij}\neq 0$ for some $i,j$ this may no longer be the case. Here, we choose three independent detuning variables as $\Delta_{13}$, $\Delta_{24}$, and $\Delta_{34}$ so that the natural frequencies can be written as
\begin{equation}
\begin{split}
	\omega_1&=\omega+\Delta_{13}+\Delta_{34}\\
	\omega_2&=\omega+\Delta_{24}\\
	\omega_3&=\omega+\Delta_{34}\\
	\omega_4&=\omega.
\end{split}
\label{eq_detunedfrequencies}
\end{equation}

Using (\ref{eq_detunedfrequencies}) instead of (\ref{eq_identity}), phase difference system (\ref{eqndphidt}) can be rewritten as
\begin{eqnarray}
\dot{\phi_1}&=&\Delta_{13}+f(\phi_1;\phi_2-\phi_3,0)
-f(0;\phi_1,\phi_2-\phi_3)\nonumber\\
\dot{\phi_2}&=&\Delta_{24}+f(\phi_2;\phi_1+\phi_3,0) -f(0;\phi_1+\phi_3,\phi_2)\label{eqndphidt_with_detunings}\\
\dot{\phi_3}&=&\Delta_{34}+f(\phi_3;\phi_1+\phi_3,\phi_2)
-f(0;\phi_1+\phi_3,\phi_2).\nonumber
\end{eqnarray}

An interesting thing about the heteroclinic ratchet (such as that illustrated in Figure \ref{fig_heteroclinic_ratchet}) is that the qualitative response to detuning depends on the sign of the detuning.
An example showing $\Omega_{13}$, the difference between the observed average frequencies of the oscillators $1$ and $3$, as a function of $\Delta_{13}$ is given in Figure~\ref{fig_detuning}. Considering (\ref{eqndphidt_with_detunings}), one can observe that since the heteroclinic ratchet includes winding connections in the $+\phi_1$ direction but no connections winding in the $-\phi_1$ direction, the oscillator system responds to $\Delta_{13}>0$ by breaking frequency synchronization of the oscillator pair $(1,3)$, whereas $\Delta_{13}\leq 0$ leaves the frequency synchronization unchanged, $\Omega_{13}=0$. There is a similar response for the difference between
oscillators $2$ and $4$ as can be seen by the symmetry of the original system. Small positive and/or negative detunings $\Delta_{34}$ do not have any qualitative effect on dynamics of (\ref{eqndphidt_with_detunings}) near the heteroclinic ratchet considered, since it does not include winding connections in the $+\phi_3$ or $-\phi_3$ directions.

\subsection{Response of the system to noise and detuning}

\begin{figure}
\begin{center}
	\includegraphics[width=0.60\textwidth]{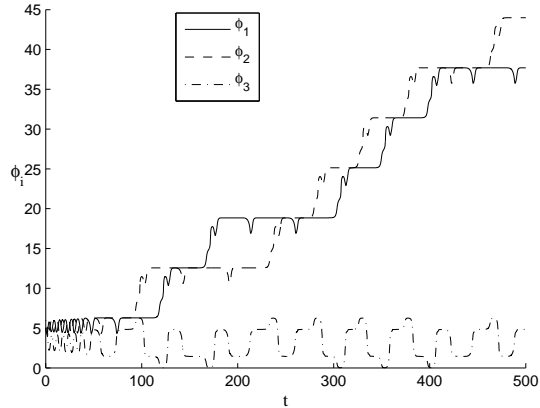}
\end{center}
	\caption{\sc A solution of the system (\ref{eqn3cellspecific}) with no detuning and additive white noise ($\text{amplitude}=10^{-6}$) for the parameter set (\ref{eq_parameterset1}).  
%The solution converges to the heteroclinic ratchet until the noise draws it to the other side of the invariant surfaces $V_3^1$ and $V_3^2$ resulting in phase slips in the $\phi_1$ and $\phi_2$ directions.
Noise causes the system to have repeated phase slips in the $+\phi_1$ and $+\phi_2$ directions.}
	\label{fig_heteroclinic_ratchet_solution}
\end{figure}

Here, we consider the effect of additive white noise with amplitude $\varepsilon$ for the system (\ref{eqndphidt_with_detunings}) with $\Delta_{34}=0$ and $\Delta_{13}=\Delta_{24}=\Delta$.
Recall that the heteroclinic cycle shown in Figure~\ref{fig_heteroclinic_ratchet} contains two nonwinding and two winding trajectories, and in the ideal case (no noise and no detuning) a solution converging to the heteroclinic ratchet oscillates near the nonwinding trajectories. However, addition of noise to the system without detuning will cause phase slips in $+\phi_1$ and $+\phi_2$ directions such that winding will be present even for arbitrary low amplitude $\varepsilon$ (see Figure~\ref{fig_heteroclinic_ratchet_solution}). 

\begin{figure}
\begin{center}
	\includegraphics[width=1\textwidth]{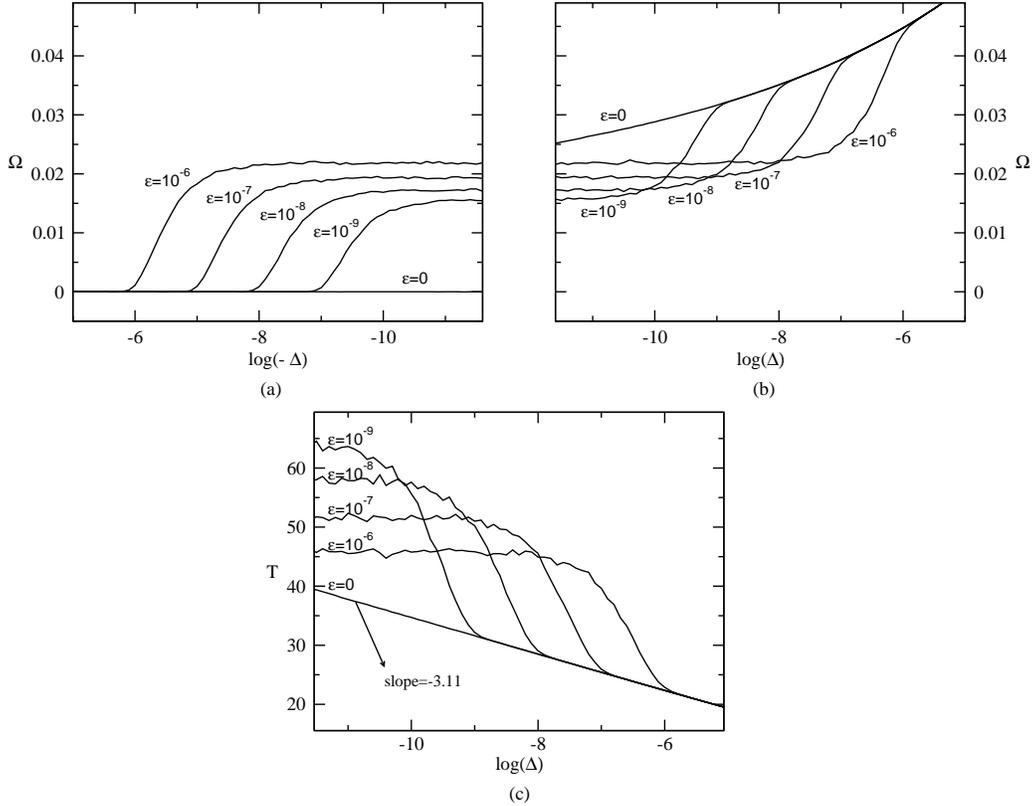}
\end{center}
	\caption{\sc Winding frequency $\Omega$ plotted against $\log\left(\Delta\right)$ for (a) $\Delta<0$, (b) $\Delta>0$ and additive noise of amplitude $\varepsilon$. The corresponding winding period $T=\Omega^{-1}$ is plotted in (c) for $\Delta>0$. Note that for $|\Delta|\ll\varepsilon$, noise dominates causing a $\Delta$-independent winding, while $\Delta>\varepsilon$ implies winding and $\Delta<-\varepsilon$ gives no winding. The winding period $T$ varies linearly with $\log\left(\Delta\right)$ until noise effects dominate.} 
	\label{fig_detunings}
\end{figure}

Let us define a winding frequency of the system (\ref{eq4cell4dimension}) as $\Omega=(\Omega_{13}+\Omega_{24})/(2\pi)$ and the corresponding winding period as $\text T=\Omega^{-1}$. For a given noise amplitude $\varepsilon$ and detuning $\Delta$, the winding frequency $\Omega\left(\varepsilon,\Delta\right)$ can be obtained numerically as in Figure~\ref{fig_detunings}. 
Even in the presence of negative detuning $\Delta<0$,
arbitrarily low amplitude noise will eventually cause fluctuations such that the winding trajectories in the ratchet are visited. 
This can be seen from Figure~\ref{fig_detunings}a, where $\Omega$ is plotted as a function of $\Delta<0$ for different noise amplitudes $\varepsilon$. 

The effect of noise on the dynamics near the heteroclinic ratchet is different when $\Delta>0$ is considered. In this case noise can cause fluctuations such that nonwinding trajectories are visited more frequently than in the case of positive detuning without noise. This happens only when $0<\Delta\ll\varepsilon$, and diminishes the observed winding frequency $\Omega$.

Note that the winding period T in the absence of noise varies linearly with $\log\left(\Delta\right)$ for $0<\Delta\ll1$ (see Figure~\ref{fig_detunings}c). It is because T can be expressed in terms of $\Delta$ as
$$\text
T(0,\Delta)=\Omega\left(0,\Delta\right)\cong-\frac{1}{\lambda}\ln\left(\Delta\right)=-\frac{\ln\left(10\right)}{\lambda}
\log\left(\Delta\right),$$ as expected from the residence time near an equilibrium of a
perturbed homoclinic cycle \cite{stone_90}, where $\lambda$ is the most positive
eigenvalue at the saddle and $\log=\log_{10}$. In our case, $\lambda=0.74$ as found in
Section~\ref{sec_ratchets} and the corresponding
slope of line representing the relation between T and $\log\left(\Delta\right)$ is $-\ln\left(10\right)/\lambda=-3.11$, consistent with simulations (see Figure~\ref{fig_detunings}c).

In the absence of detuning the winding period depends on the noise amplitude in a similar way but with a multiplier $2$, that is $\text T(\varepsilon,0)=2\: \text T(0,\varepsilon)$. In order to see this, recall that the heteroclinic ratchet contains one winding and one nonwinding trajectory from $p$ (or $q$). Since a solution converging to a heteroclinic network spends most of its time near equilibria we can consider the effect of weak noise as perturbations near the equilibria. Considering the lower (upper) equilibrium $p$ ($q$), nonwinding and winding trajectories are chosen with equal probabilities in case of the unbiased homogeneous noise as a result of the presence of invariant subspace $V_3^2$ ($V_3^1$). Therefore, on average, a trajectory in the presence of weak noise approaches both equilibria $p$ and $q$ in one winding period. Thus, the winding period is twice as large as the winding period for $\varepsilon=0$ and $\Delta>0$ where the trajectories passes one equilibria in each winding period as only the winding trajectories of the ratchet are visited. The consequence of this can also be seen in Figure~\ref{fig_detunings}b, where 
%Note that for a specific $\varepsilon$, $\Omega$ converges to a value $\Omega(\varepsilon,0)$ as $\Delta$ tends to zero, and this value is close to half of the frequency observed at $\Delta=\varepsilon$ for zero noise (see Figure \ref{fig_detunings}b), that is 
$\Omega\left(\varepsilon,\Delta\right)\cong\Omega\left(0,\varepsilon\right)/2$ for $0<\Delta\ll\varepsilon$.

% So far we have not undertaken a detailed study of the relative effects or scalings of $\Omega_{13}$ with detuning $\Delta_{13}$ and noise amplitude $\eta$, but this should be relatively easily computable, ..... {\bf MORE, maybe figure?}

\subsection{Frequency synchronization without phase synchronization}

Adding unbiased homogeneous noise (without detuning) can lead to frequency synchronization {\em without} phase synchronization; one can have a situation where $\phi_1$ and $\phi_2$ are frequency synchronized but $\phi_1-\phi_2$ is unbounded. 
%To see this, consider the heteroclinic ratchet shown in Figure \ref{fig_heteroclinic_ratchet}c. Since a solution converging to a heteroclinic network spends most of its time near equlibria we can consider the effect of weak noise as perturbations near the equilibria. Considering the lower (upper) equilibrium $p$ ($q$), 
It is because the presence of unbiased noise means that the average frequency of the phase slips in the $+\phi_1$ and $+\phi_2$ directions should be equal, that is $\lim_{t\rightarrow \infty}\frac{\phi_1-\phi_2}{t}=0$. 

Using the usual phase variables we can write this as 
\begin{equation}
\label{eq_multifrequencylocking}
	\lim_{t\rightarrow \infty}\frac{\theta_1-\theta_3-\theta_2+\theta_4}{t}=0.
\end{equation}
Due to the symmetry of the system when the detunings are zero, we have $\Omega_{34}=\lim_{t\rightarrow \infty}\frac{\phi_3}{t}=\lim_{t\rightarrow \infty}\frac{\theta_3-\theta_4}{t}=0$. Thus, (\ref{eq_multifrequencylocking}) implies $\Omega_{12}=\lim_{t\rightarrow \infty}\frac{\theta_1-\theta_2}{t}=0$, namely, the frequency synchronization of the oscillators 1 and 2.

As a result, on the heteroclinic ratchet, arbitrary small homogeneous noise will cause all oscillator pairs to lose phase synchronization. Moreover, the oscillator pairs $(1,3)$ and $(2,4)$ lose their frequency synchronization, whereas the pairs $(1,2)$ and $(3,4)$ maintain their frequency synchronization but lose phase synchronization.

% The oscillators $1$ and $2$ and the oscillators $3$ and $4$ are frequency synchronized although the phases of the former oscillator pair advance faster than the phases of the latter pair, which
% means both phase and frequency synchronizations are lost for the oscillator pairs $(1,3)$ and $(2,4)$ (see Figure \ref{fig_heteroclinic_ratchet_solution}). Note that in this case none of the pairs except the $(3,4)$ are phase synchronized since due to the noise any bound can
% be exceeded.

\subsection{Other comments}

As the existence and robustness of heteroclinic ratchets relies only on the presence of invariant subspaces and the existence of robust heteroclinic cycles, we believe that heteroclinic ratchets will be present in a variety of coupled dynamical systems. Moreover, they will not
occur in purely symmetry-forced heteroclinic networks because these will have unstable manifold branches that are symmetrically related. 

The four cell example we have discussed here is interesting in that we believe it is in some sense the simplest; for example, robust heteroclinic attractors cannot occur in fewer than four globally coupled oscillators. In applications, one can think of the network as a possible dynamical {\em motif} \cite{zhigulin_04}, i.e. a dynamical building block for network with a more complex function. Motifs in networks have been investigated in different areas since the work of Milo et.al. \cite{milo_02}, and asymmetrically coupled small networks are found to exist in neural networks as functional motifs \cite{sporns_04}. 

The analysis of the considered system in the presence of detuning shows that {\em extreme sensitivity to detuning} \cite{ashwin_burylko_maistrenko_popovych_06} may be a subtle phenomenon with for example rectification properties, and we conjecture such dynamical functions may be of use for information processing, for example in neural systems.

There remain a number of questions and details to be investigated for the example presented here; for instance, understanding the detailed dynamics on adding nonzero detuning will be quite a challenge, as will be obtaining a full understanding of the bifurcation structure.

\subsection*{Acknowledgements}

We thank Mike Field for discussions relating to this work.

\section*{Appendix: Unstability of the heteroclinic ratchet for $r_k=0,\ k=3,4,\dots$}
In Section \ref{sec_ratchets}, it is shown that for the system (\ref{eqn3cellspecific}), an asymptotically stable heteroclinic ratchet exists that connects the equilibria $p$ and $q$. Here, we show that this heteroclinic ratchet cannot be asymptotically stable if only the first two harmonics of the coupling function is considered. That is,
\begin{equation}
\label{eq_apppendix_couplingfunction}
 g(x)=-\sin (x+\alpha_1)+r_2\sin (2x+\alpha_2).
\end{equation}	
The equilibria $p=(0,0,p_3)$ and $q=(0,0,2\pi-p_3)$ in $V_2$ are given by
\begin{equation}
 p_3=\cos^{-1}\left(\frac{\cos\alpha_1}{2r_2\cos\alpha_2}\right).
\label{eq_appendix_equilibrium}
\end{equation}
This can be obtained from (\ref{eqn3cellspecific}) by setting $\phi_1=\phi_2=\dot\phi_3=0$.
Let us calculate the eigenvalues at $p$. Linearising (\ref{eqn3cellspecific}) at $p$ gives 
\begin{equation}
 \lambda_{1,2}=g'(\mp p)+2g'(0),\ \lambda_3=g'(p)+g'(-p),
\label{eq_lambdas}
\end{equation}
where $\lambda_1$ and $\lambda_2$ corresponds to the eigenvectors in
$V_3^1$ and $V_3^2$ respectively, and $\lambda_3$ is the radial eigenvalue which corresponds to the eigenvector in $V_2=V_3^1\cap V_3^2$.

For the existence of an asymptotically stable heteroclinic network connecting the equilibrium $p$ to its symmetric image $q=\rho(p)$ the following conditions are necessary:
\begin{align}
	&\text{Existence of nodes $p$ and $q$: } \left|\frac{\cos\alpha_1}{2r_2\cos\alpha_2}\right|<1\label{eq_appendix_existence}\\
&\text{Negative radial eigenvalue: } \lambda_3=g'(p)+g'(-p)<0\label{eq_appendix_radial}\\
&\text{Asymptotic stability condition: }\lambda_1+\lambda_2=g'(p)+g'(-p)+4g'(0)<0\label{eq_appendix_stability}
\end{align}

We first assume $r_2\cos\alpha_2<0$. From (\ref{eq_appendix_radial}) we have
\begin{align}
	&-2\cos p\cos\alpha_1+4r_2\cos 2p\cos\alpha_2<0\\
	&-2\cos p\cos\alpha_1+8r_2\cos^2p\cos\alpha_2-4r_2\cos\alpha_2<0.
\end{align}
Substituting (\ref{eq_appendix_equilibrium}) we get
\begin{equation*}
	\frac{\cos^2\alpha_1}{r_2\cos\alpha_2}-4r_2\cos\alpha_2<0.
\end{equation*}
Our assumption then follows
\begin{equation*}
	\frac{\cos^2\alpha_1}{4r_2^2\cos^2\alpha_2}>1,
\end{equation*}
which contradicts (\ref{eq_appendix_existence}). On the other hand, if we assume $r_2\cos\alpha_2>0$, the condition (\ref{eq_appendix_stability}) cannot be satisfied since
\begin{eqnarray}
 \lambda_1+\lambda_2&=&g'(p)+g'(-p)+4g'(0)\nonumber\\
 &=&-2\cos p\cos\alpha_1+8r_2\cos^2p\cos\alpha_2-4r_2\cos\alpha_2-4\cos\alpha_1+8r_2\cos\alpha_2\nonumber
\end{eqnarray}
and substituting (\ref{eq_appendix_equilibrium}) one gets
\begin{eqnarray*}
 \lambda_1+\lambda_2&=&-\frac{\cos^2\alpha_1}{r_2\cos\alpha_2}+\frac{2\cos^2\alpha_1}{r_2\cos\alpha_2}+4r_2\cos\alpha_2-4\cos\alpha_1\\
 &=&\left(\frac{\cos\alpha_1}{\sqrt{r_2\cos\alpha_2}}-2\sqrt{r_2\cos\alpha_2}\right)^2\geq0.
\label{eq_sumoflambdas}
\end{eqnarray*}
Thus, the expanding eigenvalue is greater in
absolute value than the contracting eigenvalue.

\bibliographystyle{plain} 
\bibliography{4-Cell.bib}

\end{document}

%% file: 4cell.pstex_t
\begin{picture}(0,0)%
\includegraphics{4cell.pstex}%
\end{picture}%
\setlength{\unitlength}{3947sp}%
\begingroup\makeatletter\ifx\SetFigFont\undefined%
\gdef\SetFigFont#1#2#3#4#5{%
  \reset@font\fontsize{#1}{#2pt}%
  \fontfamily{#3}\fontseries{#4}\fontshape{#5}%
  \selectfont}%
\fi\endgroup%
\begin{picture}(2199,1731)(364,-1255)
\put(2296, 89){\makebox(0,0)[lb]{\smash{{\SetFigFont{12}{14.4}{\familydefault}{\mddefault}{\updefault}{\color[rgb]{0,0,0}2}%
}}}}
\put(2296,-1096){\makebox(0,0)[lb]{\smash{{\SetFigFont{12}{14.4}{\familydefault}{\mddefault}{\updefault}{\color[rgb]{0,0,0}4}%
}}}}
\put(556,-1111){\makebox(0,0)[lb]{\smash{{\SetFigFont{12}{14.4}{\familydefault}{\mddefault}{\updefault}{\color[rgb]{0,0,0}3}%
}}}}
\put(571, 89){\makebox(0,0)[lb]{\smash{{\SetFigFont{12}{14.4}{\familydefault}{\mddefault}{\updefault}{\color[rgb]{0,0,0}1}%
}}}}
\end{picture}%

%% file: invariantsubsets.pstex_t
\begin{picture}(0,0)%
\includegraphics{invariantsubsets.pstex}%
\end{picture}%
\setlength{\unitlength}{2763sp}%
\begingroup\makeatletter\ifx\SetFigFontNFSS\undefined%
\gdef\SetFigFontNFSS#1#2#3#4#5{%
  \reset@font\fontsize{#1}{#2pt}%
  \fontfamily{#3}\fontseries{#4}\fontshape{#5}%
  \selectfont}%
\fi\endgroup%
\begin{picture}(6180,5176)(1786,-5720)
\put(5851,-5611){\makebox(0,0)[lb]{\smash{{\SetFigFontNFSS{14}{16.8}{\familydefault}{\mddefault}{\updefault}{\color[rgb]{0,0,0}$V^{s3}_2$}%
}}}}
\put(4351,-811){\makebox(0,0)[lb]{\smash{{\SetFigFontNFSS{14}{16.8}{\familydefault}{\mddefault}{\updefault}{\color[rgb]{0,0,0}${V_2}$}%
}}}}
\put(1801,-5461){\makebox(0,0)[lb]{\smash{{\SetFigFontNFSS{14}{16.8}{\familydefault}{\mddefault}{\updefault}{\color[rgb]{0,0,0}$V^{s1}_2$}%
}}}}
\put(7951,-4111){\makebox(0,0)[lb]{\smash{{\SetFigFontNFSS{14}{16.8}{\familydefault}{\mddefault}{\updefault}{\color[rgb]{0,0,0}$V^{s2}_2$}%
}}}}
\put(5401,-2911){\makebox(0,0)[lb]{\smash{{\SetFigFontNFSS{14}{16.8}{\familydefault}{\mddefault}{\updefault}{\color[rgb]{0,0,0}$V_3^2$}%
}}}}
\put(4126,-5536){\makebox(0,0)[lb]{\smash{{\SetFigFontNFSS{14}{16.8}{\familydefault}{\mddefault}{\updefault}{\color[rgb]{0,0,0}$V^s_3$}%
}}}}
\put(5926,-4336){\makebox(0,0)[lb]{\smash{{\SetFigFontNFSS{14}{16.8}{\familydefault}{\mddefault}{\updefault}{\color[rgb]{0,0,0}$\bar V^{s3}_2$}%
}}}}
\put(3301,-3361){\rotatebox{25.0}{\makebox(0,0)[lb]{\smash{{\SetFigFontNFSS{14}{16.8}{\familydefault}{\mddefault}{\updefault}{\color[rgb]{0,0,0}$V_3^1$}%
}}}}}
\put(4201,-1561){\makebox(0,0)[lb]{\smash{{\SetFigFontNFSS{10}{12.0}{\familydefault}{\mddefault}{\updefault}{\color[rgb]{0,0,0}$2\pi$}%
}}}}
\put(2401,-4861){\makebox(0,0)[lb]{\smash{{\SetFigFontNFSS{10}{12.0}{\familydefault}{\mddefault}{\updefault}{\color[rgb]{0,0,0}$2\pi$}%
}}}}
\put(4276,-3961){\makebox(0,0)[lb]{\smash{{\SetFigFontNFSS{10}{12.0}{\familydefault}{\mddefault}{\updefault}{\color[rgb]{0,0,0}$0$}%
}}}}
\put(6976,-3886){\makebox(0,0)[lb]{\smash{{\SetFigFontNFSS{10}{12.0}{\familydefault}{\mddefault}{\updefault}{\color[rgb]{0,0,0}$2\pi$}%
}}}}
\end{picture}%

%% file: 3cellV3a.pstex_t
\begin{picture}(0,0)%
\includegraphics{3cellV3a.pstex}%
\end{picture}%
\setlength{\unitlength}{1973sp}%
\begingroup\makeatletter\ifx\SetFigFont\undefined%
\gdef\SetFigFont#1#2#3#4#5{%
  \reset@font\fontsize{#1}{#2pt}%
  \fontfamily{#3}\fontseries{#4}\fontshape{#5}%
  \selectfont}%
\fi\endgroup%
\begin{picture}(3953,2839)(226,-3344)
\put(226,-1936){\makebox(0,0)[lb]{\smash{{\SetFigFont{12}{14.4}{\familydefault}{\mddefault}{\updefault}{\color[rgb]{0,0,0}$N_1$:}%
}}}}
\put(914,-661){\makebox(0,0)[lb]{\smash{{\SetFigFont{6}{7.2}{\familydefault}{\mddefault}{\updefault}{\color[rgb]{0,0,0}$\theta_1$}%
}}}}
\put(4126,-661){\makebox(0,0)[lb]{\smash{{\SetFigFont{6}{7.2}{\familydefault}{\mddefault}{\updefault}{\color[rgb]{0,0,0}$\theta_2$}%
}}}}
\put(2326,-3286){\makebox(0,0)[lb]{\smash{{\SetFigFont{6}{7.2}{\familydefault}{\mddefault}{\updefault}{\color[rgb]{0,0,0}$\theta_3=\theta_4$}%
}}}}
\end{picture}%

%% file: 3cellV3b2.pstex_t
\begin{picture}(0,0)%
\includegraphics{3cellV3b2.pstex}%
\end{picture}%
\setlength{\unitlength}{1973sp}%
\begingroup\makeatletter\ifx\SetFigFontNFSS\undefined%
\gdef\SetFigFontNFSS#1#2#3#4#5{%
  \reset@font\fontsize{#1}{#2pt}%
  \fontfamily{#3}\fontseries{#4}\fontshape{#5}%
  \selectfont}%
\fi\endgroup%
\begin{picture}(4227,2952)(-89,-3439)
\put(-74,-2011){\makebox(0,0)[lb]{\smash{{\SetFigFontNFSS{12}{14.4}{\familydefault}{\mddefault}{\updefault}{\color[rgb]{0,0,0}$N_2$:}%
}}}}
\put(574,-676){\makebox(0,0)[lb]{\smash{{\SetFigFontNFSS{6}{7.2}{\familydefault}{\mddefault}{\updefault}{\color[rgb]{0,0,0}$\theta_1$}%
}}}}
\put(3376,-1561){\makebox(0,0)[lb]{\smash{{\SetFigFontNFSS{6}{7.2}{\familydefault}{\mddefault}{\updefault}{\color[rgb]{0,0,0}$\theta_2=\theta_4$}%
}}}}
\put(2101,-3361){\makebox(0,0)[lb]{\smash{{\SetFigFontNFSS{6}{7.2}{\familydefault}{\mddefault}{\updefault}{\color[rgb]{0,0,0}$\theta_3$}%
}}}}
\end{picture}%

%% file: 3cellV3b1.pstex_t
\begin{picture}(0,0)%
\includegraphics{3cellV3b1.pstex}%
\end{picture}%
\setlength{\unitlength}{1973sp}%
\begingroup\makeatletter\ifx\SetFigFontNFSS\undefined%
\gdef\SetFigFontNFSS#1#2#3#4#5{%
  \reset@font\fontsize{#1}{#2pt}%
  \fontfamily{#3}\fontseries{#4}\fontshape{#5}%
  \selectfont}%
\fi\endgroup%
\begin{picture}(4043,2952)(136,-3364)
\put(151,-2011){\makebox(0,0)[lb]{\smash{{\SetFigFontNFSS{12}{14.4}{\familydefault}{\mddefault}{\updefault}{\color[rgb]{0,0,0}$N_3$:}%
}}}}
\put(4071,-641){\makebox(0,0)[lb]{\smash{{\SetFigFontNFSS{6}{7.2}{\familydefault}{\mddefault}{\updefault}{\color[rgb]{0,0,0}$\theta_2$}%
}}}}
\put(828,-1411){\makebox(0,0)[lb]{\smash{{\SetFigFontNFSS{6}{7.2}{\familydefault}{\mddefault}{\updefault}{\color[rgb]{0,0,0}$\theta_1=\theta_3$}%
}}}}
\put(2626,-3286){\makebox(0,0)[lb]{\smash{{\SetFigFontNFSS{6}{7.2}{\familydefault}{\mddefault}{\updefault}{\color[rgb]{0,0,0}$\theta_4$}%
}}}}
\end{picture}%

%% file: heteroclinic_ratchet_all_in_one.pstex_t
\begin{picture}(0,0)%
\includegraphics{heteroclinic_ratchet_all_in_one.pstex}%
\end{picture}%
\setlength{\unitlength}{1776sp}%
\begingroup\makeatletter\ifx\SetFigFontNFSS\undefined%
\gdef\SetFigFontNFSS#1#2#3#4#5{%
  \reset@font\fontsize{#1}{#2pt}%
  \fontfamily{#3}\fontseries{#4}\fontshape{#5}%
  \selectfont}%
\fi\endgroup%
\begin{picture}(12772,13836)(286,-13763)
\put(301,-1111){\makebox(0,0)[lb]{\smash{{\SetFigFontNFSS{9}{10.8}{\familydefault}{\mddefault}{\updefault}{\color[rgb]{0,0,0}$q$}%
}}}}
\put(5626,-1111){\makebox(0,0)[lb]{\smash{{\SetFigFontNFSS{9}{10.8}{\familydefault}{\mddefault}{\updefault}{\color[rgb]{0,0,0}$q$}%
}}}}
\put(301,-4111){\makebox(0,0)[lb]{\smash{{\SetFigFontNFSS{9}{10.8}{\familydefault}{\mddefault}{\updefault}{\color[rgb]{0,0,0}$p$}%
}}}}
\put(5626,-4111){\makebox(0,0)[lb]{\smash{{\SetFigFontNFSS{9}{10.8}{\familydefault}{\mddefault}{\updefault}{\color[rgb]{0,0,0}$p$}%
}}}}
\put(11410,-8707){\makebox(0,0)[lb]{\smash{{\SetFigFontNFSS{10}{12.0}{\familydefault}{\mddefault}{\updefault}{\color[rgb]{0,0,0}$\phi_2$}%
}}}}
\put(2626,-12847){\makebox(0,0)[lb]{\smash{{\SetFigFontNFSS{10}{12.0}{\familydefault}{\mddefault}{\updefault}{\color[rgb]{0,0,0}$\phi_1$}%
}}}}
\put(3881,-6223){\makebox(0,0)[lb]{\smash{{\SetFigFontNFSS{10}{12.0}{\familydefault}{\mddefault}{\updefault}{\color[rgb]{0,0,0}$\phi_3$}%
}}}}
\put(5776,-13636){\makebox(0,0)[lb]{\smash{{\SetFigFontNFSS{11}{13.2}{\familydefault}{\mddefault}{\updefault}{\color[rgb]{0,0,0}(c)}%
}}}}
\put(9299,-230){\makebox(0,0)[lb]{\smash{{\SetFigFontNFSS{10}{12.0}{\familydefault}{\mddefault}{\updefault}{\color[rgb]{0,0,0}$\phi_3$}%
}}}}
\put(6451,-5113){\makebox(0,0)[lb]{\smash{{\SetFigFontNFSS{10}{12.0}{\familydefault}{\mddefault}{\updefault}{\color[rgb]{0,0,0}$\phi_1$}%
}}}}
\put(13043,-3567){\makebox(0,0)[lb]{\smash{{\SetFigFontNFSS{10}{12.0}{\familydefault}{\mddefault}{\updefault}{\color[rgb]{0,0,0}$\phi_2$}%
}}}}
\put(7102,-2021){\makebox(0,0)[lb]{\smash{{\SetFigFontNFSS{9}{10.8}{\familydefault}{\mddefault}{\updefault}{\color[rgb]{0,0,0}$q$}%
}}}}
\put(7102,-4299){\makebox(0,0)[lb]{\smash{{\SetFigFontNFSS{9}{10.8}{\familydefault}{\mddefault}{\updefault}{\color[rgb]{0,0,0}$p$}%
}}}}
\put(12148,-3323){\makebox(0,0)[lb]{\smash{{\SetFigFontNFSS{9}{10.8}{\familydefault}{\mddefault}{\updefault}{\color[rgb]{0,0,0}$p$}%
}}}}
\put(12148,-1369){\makebox(0,0)[lb]{\smash{{\SetFigFontNFSS{9}{10.8}{\familydefault}{\mddefault}{\updefault}{\color[rgb]{0,0,0}$q$}%
}}}}
\put(9478,-1500){\makebox(0,0)[lb]{\smash{{\SetFigFontNFSS{9}{10.8}{\familydefault}{\mddefault}{\updefault}{\color[rgb]{0,0,0}$q$}%
}}}}
\put(10105,-1949){\makebox(0,0)[lb]{\smash{{\SetFigFontNFSS{9}{10.8}{\familydefault}{\mddefault}{\updefault}{\color[rgb]{0,0,0}$q$}%
}}}}
\put(10502,-4353){\makebox(0,0)[lb]{\smash{{\SetFigFontNFSS{9}{10.8}{\familydefault}{\mddefault}{\updefault}{\color[rgb]{0,0,0}$p$}%
}}}}
\put(9462,-3567){\makebox(0,0)[lb]{\smash{{\SetFigFontNFSS{9}{10.8}{\familydefault}{\mddefault}{\updefault}{\color[rgb]{0,0,0}$p$}%
}}}}
\put(11476,-511){\makebox(0,0)[lb]{\smash{{\SetFigFontNFSS{10}{12.0}{\familydefault}{\mddefault}{\updefault}{\color[rgb]{0,0,0}$V_3^2$}%
}}}}
\put(8176,-661){\makebox(0,0)[lb]{\smash{{\SetFigFontNFSS{10}{12.0}{\familydefault}{\mddefault}{\updefault}{\color[rgb]{0,0,0}$V_3^1$}%
}}}}
\put(9462,-5730){\makebox(0,0)[lb]{\smash{{\SetFigFontNFSS{11}{13.2}{\familydefault}{\mddefault}{\updefault}{\color[rgb]{0,0,0}(b)}%
}}}}
\put(2716,-5791){\makebox(0,0)[lb]{\smash{{\SetFigFontNFSS{11}{13.2}{\familydefault}{\mddefault}{\updefault}{\color[rgb]{0,0,0}(a)}%
}}}}
\end{picture}%

%% file: bifurcating_ratchet_all_in_one.pstex_t
\begin{picture}(0,0)%
\includegraphics{./bifurcating_ratchet_all_in_one.pstex}%
\end{picture}%
\setlength{\unitlength}{1381sp}%
\begingroup\makeatletter\ifx\SetFigFont\undefined%
\gdef\SetFigFont#1#2#3#4#5{%
  \reset@font\fontsize{#1}{#2pt}%
  \fontfamily{#3}\fontseries{#4}\fontshape{#5}%
  \selectfont}%
\fi\endgroup%
\begin{picture}(17140,12591)(451,-12460)
\put(11851,-1186){\makebox(0,0)[lb]{\smash{{\SetFigFont{6}{7.2}{\familydefault}{\mddefault}{\updefault}{\color[rgb]{0,0,0}$q$}%
}}}}
\put(17176,-1186){\makebox(0,0)[lb]{\smash{{\SetFigFont{6}{7.2}{\familydefault}{\mddefault}{\updefault}{\color[rgb]{0,0,0}$q$}%
}}}}
\put(11851,-4186){\makebox(0,0)[lb]{\smash{{\SetFigFont{6}{7.2}{\familydefault}{\mddefault}{\updefault}{\color[rgb]{0,0,0}$p$}%
}}}}
\put(17176,-4186){\makebox(0,0)[lb]{\smash{{\SetFigFont{6}{7.2}{\familydefault}{\mddefault}{\updefault}{\color[rgb]{0,0,0}$p$}%
}}}}
\put(11476,-4186){\makebox(0,0)[lb]{\smash{{\SetFigFont{6}{7.2}{\familydefault}{\mddefault}{\updefault}{\color[rgb]{0,0,0}$p$}%
}}}}
\put(11476,-1186){\makebox(0,0)[lb]{\smash{{\SetFigFont{6}{7.2}{\familydefault}{\mddefault}{\updefault}{\color[rgb]{0,0,0}$q$}%
}}}}
\put(6151,-1186){\makebox(0,0)[lb]{\smash{{\SetFigFont{6}{7.2}{\familydefault}{\mddefault}{\updefault}{\color[rgb]{0,0,0}$q$}%
}}}}
\put(6151,-4186){\makebox(0,0)[lb]{\smash{{\SetFigFont{6}{7.2}{\familydefault}{\mddefault}{\updefault}{\color[rgb]{0,0,0}$p$}%
}}}}
\put(5776,-4186){\makebox(0,0)[lb]{\smash{{\SetFigFont{6}{7.2}{\familydefault}{\mddefault}{\updefault}{\color[rgb]{0,0,0}$p$}%
}}}}
\put(5776,-1186){\makebox(0,0)[lb]{\smash{{\SetFigFont{6}{7.2}{\familydefault}{\mddefault}{\updefault}{\color[rgb]{0,0,0}$q$}%
}}}}
\put(451,-4186){\makebox(0,0)[lb]{\smash{{\SetFigFont{6}{7.2}{\familydefault}{\mddefault}{\updefault}{\color[rgb]{0,0,0}$p$}%
}}}}
\put(1201,-5236){\makebox(0,0)[lb]{\smash{{\SetFigFont{6}{7.2}{\familydefault}{\mddefault}{\updefault}{\color[rgb]{0,0,0}$s$}%
}}}}
\put(1201,-61){\makebox(0,0)[lb]{\smash{{\SetFigFont{6}{7.2}{\familydefault}{\mddefault}{\updefault}{\color[rgb]{0,0,0}$s$}%
}}}}
\put(451,-1186){\makebox(0,0)[lb]{\smash{{\SetFigFont{6}{7.2}{\familydefault}{\mddefault}{\updefault}{\color[rgb]{0,0,0}$q$}%
}}}}
\put(3792,-6828){\makebox(0,0)[lb]{\smash{{\SetFigFont{8}{9.6}{\familydefault}{\mddefault}{\updefault}{\color[rgb]{0,0,0}$\phi_3$}%
}}}}
\put(969,-11667){\makebox(0,0)[lb]{\smash{{\SetFigFont{8}{9.6}{\familydefault}{\mddefault}{\updefault}{\color[rgb]{0,0,0}$\phi_1$}%
}}}}
\put(7502,-10135){\makebox(0,0)[lb]{\smash{{\SetFigFont{8}{9.6}{\familydefault}{\mddefault}{\updefault}{\color[rgb]{0,0,0}$\phi_2$}%
}}}}
\put(1614,-8602){\makebox(0,0)[lb]{\smash{{\SetFigFont{7}{8.4}{\familydefault}{\mddefault}{\updefault}{\color[rgb]{0,0,0}$q$}%
}}}}
\put(1614,-10861){\makebox(0,0)[lb]{\smash{{\SetFigFont{7}{8.4}{\familydefault}{\mddefault}{\updefault}{\color[rgb]{0,0,0}$p$}%
}}}}
\put(6615,-9893){\makebox(0,0)[lb]{\smash{{\SetFigFont{7}{8.4}{\familydefault}{\mddefault}{\updefault}{\color[rgb]{0,0,0}$p$}%
}}}}
\put(5002,-10861){\makebox(0,0)[lb]{\smash{{\SetFigFont{7}{8.4}{\familydefault}{\mddefault}{\updefault}{\color[rgb]{0,0,0}$p$}%
}}}}
\put(6615,-7957){\makebox(0,0)[lb]{\smash{{\SetFigFont{7}{8.4}{\familydefault}{\mddefault}{\updefault}{\color[rgb]{0,0,0}$q$}%
}}}}
\put(5002,-8602){\makebox(0,0)[lb]{\smash{{\SetFigFont{7}{8.4}{\familydefault}{\mddefault}{\updefault}{\color[rgb]{0,0,0}$q$}%
}}}}
\put(6212,-10619){\makebox(0,0)[lb]{\smash{{\SetFigFont{7}{8.4}{\familydefault}{\mddefault}{\updefault}{\color[rgb]{0,0,0}$s$}%
}}}}
\put(5688,-7753){\makebox(0,0)[lb]{\smash{{\SetFigFont{7}{8.4}{\familydefault}{\mddefault}{\updefault}{\color[rgb]{0,0,0}$s$}%
}}}}
\put(3198,-7710){\makebox(0,0)[lb]{\smash{{\SetFigFont{7}{8.4}{\familydefault}{\mddefault}{\updefault}{\color[rgb]{0,0,0}$s$}%
}}}}
\put(4437,-7554){\makebox(0,0)[lb]{\smash{{\SetFigFont{7}{8.4}{\familydefault}{\mddefault}{\updefault}{\color[rgb]{0,0,0}$s$}%
}}}}
\put(2582,-11425){\makebox(0,0)[lb]{\smash{{\SetFigFont{7}{8.4}{\familydefault}{\mddefault}{\updefault}{\color[rgb]{0,0,0}$s$}%
}}}}
\put(2582,-8199){\makebox(0,0)[lb]{\smash{{\SetFigFont{7}{8.4}{\familydefault}{\mddefault}{\updefault}{\color[rgb]{0,0,0}$s$}%
}}}}
\put(3593,-8140){\makebox(0,0)[lb]{\smash{{\SetFigFont{7}{8.4}{\familydefault}{\mddefault}{\updefault}{\color[rgb]{0,0,0}$q$}%
}}}}
\put(4437,-10135){\makebox(0,0)[lb]{\smash{{\SetFigFont{7}{8.4}{\familydefault}{\mddefault}{\updefault}{\color[rgb]{0,0,0}$s$}%
}}}}
\put(3550,-10619){\makebox(0,0)[lb]{\smash{{\SetFigFont{7}{8.4}{\familydefault}{\mddefault}{\updefault}{\color[rgb]{0,0,0}$s$}%
}}}}
\put(3615,-9796){\makebox(0,0)[lb]{\smash{{\SetFigFont{7}{8.4}{\familydefault}{\mddefault}{\updefault}{\color[rgb]{0,0,0}$p$}%
}}}}
\put(12665,-6887){\makebox(0,0)[lb]{\smash{{\SetFigFont{8}{9.6}{\familydefault}{\mddefault}{\updefault}{\color[rgb]{0,0,0}$\phi_3$}%
}}}}
\put(9751,-11740){\makebox(0,0)[lb]{\smash{{\SetFigFont{8}{9.6}{\familydefault}{\mddefault}{\updefault}{\color[rgb]{0,0,0}$\phi_1$}%
}}}}
\put(16493,-10204){\makebox(0,0)[lb]{\smash{{\SetFigFont{8}{9.6}{\familydefault}{\mddefault}{\updefault}{\color[rgb]{0,0,0}$\phi_2$}%
}}}}
\put(10417,-8667){\makebox(0,0)[lb]{\smash{{\SetFigFont{7}{8.4}{\familydefault}{\mddefault}{\updefault}{\color[rgb]{0,0,0}$q$}%
}}}}
\put(10417,-10931){\makebox(0,0)[lb]{\smash{{\SetFigFont{7}{8.4}{\familydefault}{\mddefault}{\updefault}{\color[rgb]{0,0,0}$p$}%
}}}}
\put(15577,-9961){\makebox(0,0)[lb]{\smash{{\SetFigFont{7}{8.4}{\familydefault}{\mddefault}{\updefault}{\color[rgb]{0,0,0}$p$}%
}}}}
\put(15577,-8019){\makebox(0,0)[lb]{\smash{{\SetFigFont{7}{8.4}{\familydefault}{\mddefault}{\updefault}{\color[rgb]{0,0,0}$q$}%
}}}}
\put(13488,-8596){\makebox(0,0)[lb]{\smash{{\SetFigFont{7}{8.4}{\familydefault}{\mddefault}{\updefault}{\color[rgb]{0,0,0}$q$}%
}}}}
\put(13894,-10986){\makebox(0,0)[lb]{\smash{{\SetFigFont{7}{8.4}{\familydefault}{\mddefault}{\updefault}{\color[rgb]{0,0,0}$p$}%
}}}}
\put(12847,-8181){\makebox(0,0)[lb]{\smash{{\SetFigFont{7}{8.4}{\familydefault}{\mddefault}{\updefault}{\color[rgb]{0,0,0}$q$}%
}}}}
\put(12789,-10204){\makebox(0,0)[lb]{\smash{{\SetFigFont{7}{8.4}{\familydefault}{\mddefault}{\updefault}{\color[rgb]{0,0,0}$p$}%
}}}}
\put(3001,-5761){\makebox(0,0)[lb]{\smash{{\SetFigFont{9}{10.8}{\familydefault}{\mddefault}{\updefault}{\color[rgb]{0,0,0}(a)}%
}}}}
\put(8701,-5761){\makebox(0,0)[lb]{\smash{{\SetFigFont{9}{10.8}{\familydefault}{\mddefault}{\updefault}{\color[rgb]{0,0,0}(b)}%
}}}}
\put(14401,-5761){\makebox(0,0)[lb]{\smash{{\SetFigFont{9}{10.8}{\familydefault}{\mddefault}{\updefault}{\color[rgb]{0,0,0}(c)}%
}}}}
\put(3751,-12361){\makebox(0,0)[lb]{\smash{{\SetFigFont{9}{10.8}{\familydefault}{\mddefault}{\updefault}{\color[rgb]{0,0,0}(d)}%
}}}}
\put(12601,-12361){\makebox(0,0)[lb]{\smash{{\SetFigFont{9}{10.8}{\familydefault}{\mddefault}{\updefault}{\color[rgb]{0,0,0}(e)}%
}}}}
\end{picture}%

%% file: article_4cell_5_11_08.bbl
\begin{thebibliography}{10}

\bibitem{aguiar_ashwin_dias_field_08}
M.A.D. Aguiar, P.~Ashwin, A.P.S. Dias, and M.~Field.
\newblock Robust heteroclinic cycles in coupled cell systems: identical cells
  with asymmetric inputs.
\newblock {\em preprint}, 2008.

\bibitem{aguiar_dias_golubitsky_leite_07}
M.A.D. Aguiar, A.P.S. Dias, M.~Golubitsky, and M.C.A. Leite.
\newblock Homogenous coupled cell networks with ${S}_3$-symmetric quotient.
\newblock {\em Disc. and Cont. Dyn. Sys. Supplement}, pages 1--9, 2007.

\bibitem{ashwin_borresen_04}
P.~Ashwin and J.~Borresen.
\newblock Encoding via conjugate symmetries of slow oscillations for globally
  coupled oscillators.
\newblock {\em Phy. Rev. E}, 70(2):026203, 2004.

\bibitem{ashwin_borresen_05}
P.~Ashwin and J.~Borresen.
\newblock Discrete computation using a perturbed heteroclinic network.
\newblock {\em Phy. Lett. A}, 347(4-6):208--214, 2005.

\bibitem{ashwin_burylko_maistrenko_08}
P.~Ashwin, O.~Burylko, and Y.~Maistrenko.
\newblock Bifurcation to heteroclinic cycles and sensitivity in three and four
  coupled phase oscillators.
\newblock {\em Phys. D: Non. Phe.}, 237:454--466, 2008.

\bibitem{ashwin_burylko_maistrenko_popovych_06}
P.~Ashwin, O.~Burylko, Y.~Maistrenko, and O.~Popovych.
\newblock Extreme sensitivity to detuning for globally coupled phase
  oscillators.
\newblock {\em Phy. Rev. Let.}, 96(5):054102, 2006.

\bibitem{ashwin_orosz_borresen_08}
P.~Ashwin, G.~Orosz, and J.~Borresen.
\newblock Designing the dynamics of globally coupled oscillators.
\newblock {\em preprint}, 2008.

\bibitem{ashwin_gabor_07}
P.~Ashwin, G.~Orosz, J.~Wordsworth, and S.~Townley.
\newblock Dynamics on networks of clustered states for globally coupled phase
  oscillators.
\newblock {\em SIAM J. Appl. Dyn. Sys.}, 6(4):728--758, 2007.

\bibitem{ashwin_swift_92}
P.~Ashwin and J.W. Swift.
\newblock The dynamics of {$n$} weakly coupled identical oscillators.
\newblock {\em J. Nonlinear Sci.}, 2(1):69--108, 1992.

\bibitem{busse_79}
F.H. Busse and R.M. Clever.
\newblock Nonstationary convection in a rotating system.
\newblock In U.~Müller, K.G. Roesner, and B.~Schmidt, editors, {\em Recent
  Developments in Theoretical and Experimental Fluid Dynamics}, pages 376--385.
  Springer Verlag, Berlin, 1979.

\bibitem{ermentrout}
G.B. Ermentrout.
\newblock {\em A Guide to XPPAUT for Researchers and Students}.
\newblock SIAM, Pittsburgh, 2002.

\bibitem{golubitsky_stewart}
M.~Golubitsky and I.~Stewart.
\newblock {\em The Symmetry Perspective}.
\newblock Birkh\"{a}user Verlag, Basel, 2002.

\bibitem{golubitsky_stewart_06}
M.~Golubitsky and I.~Stewart.
\newblock Nonlinear dynamics of networks: the groupoid formalism.
\newblock {\em Bull. Amer. Math. Soc. (N.S.)}, 43(3):305--364 (electronic),
  2006.

\bibitem{guckenheimer_holmes_88}
J.~Guckenheimer and P.~Holmes.
\newblock Structurally stable heteroclinic cycles.
\newblock {\em Math. Proc. Camb. Phil. Soc.}, 103:189--192, 1988.

\bibitem{hansel_mato_meunier_93}
D.~Hansel, G.~Mato, and C.~Meunier.
\newblock Clustering and slow switching in globally couppled phase oscillators.
\newblock {\em Phy. Rev. E}, 48(5):3470--3477, 1993.

\bibitem{hofbauer_sigmund_98}
J.~Hofbauer and K.~Sigmund.
\newblock {\em Evolutionary Games and Population Dynamics}.
\newblock Cambridge University Press, Cambridge, 1998.

\bibitem{kiss_07}
I.Z. Kiss, C.G. Rusin, H.~Kori, and J.L. Hudson.
\newblock Engineering complex dynamical structures: sequential patterns and
  desynchronization.
\newblock {\em Science}, 316:1886--1889, 2007.

\bibitem{kori_kuramoto_01}
H.~Kori and Y.~Kuramoto.
\newblock Slow switching in globally coupled oscillators: robustness and
  occurence through delayed coupling.
\newblock {\em Phy. Rev. E}, 63(046214), 2001.

\bibitem{krupa_97}
M.~Krupa.
\newblock Robust heteroclinic cycles.
\newblock {\em J. Nonlinear Sci.}, 7(2):129--176, 1997.

\bibitem{krupa_melbourne_95}
M.~Krupa and I.~Melbourne.
\newblock Asymptotic stability of heteroclinic cycles in systemsd with
  symmetry.
\newblock {\em Erg. Th. Dyn. Sys.}, 15:121--147, 1995.

\bibitem{kuramoto}
Y.~Kuramoto.
\newblock {\em Chemical Oscillations, Waves and Turbulence}.
\newblock Springer-Verlag, Berlin, 1984.

\bibitem{milo_02}
R.~Milo, S.~Shen-Orr, S.~Itzkovitz, N.~Kashtan, D.~Chklovskii, and U.~Alon.
\newblock Network motifs: simple building blocks of complex networks.
\newblock 298:824--827, 2002.

\bibitem{rabinovich_06}
M.I. Rabinovich, R.~Huerta, P.~Varona, and V.S. Afraimovich.
\newblock Generation and reshaping of sequences in neural systems.
\newblock {\em Bio. Cyb.}, 95:519--536, 2006.

\bibitem{rabinovich_review_06}
M.I. Rabinovich, P.~Varona, A.I. Selverston, and H.D.I. Abarbanel.
\newblock Dynamical principles in neuroscience.
\newblock {\em Rev. Mod. Phy.}, 95:519--536, 2006.

\bibitem{sakaguchi_kuramoto_86}
H.~Sakaguchi and Y.~Kuramoto.
\newblock A soluble active rotator model showing phase transitions via mutual
  entrainment.
\newblock {\em Progr. Theoret. Phys.}, 76(3):576--581, 1986.

\bibitem{sporns_04}
O~Sporns and R.~Kötter.
\newblock Motifs in brain networks.
\newblock {\em PLoS Biol}, 2(11):1910--1918, 2004.

\bibitem{stone_90}
Emily Stone and Philip Holmes.
\newblock Random perturbations of heteroclinic attractors.
\newblock {\em SIAM J. Appl. Math.}, 50(3):726--743, 1990.

\bibitem{strogatz_00}
S.H. Strogatz.
\newblock From {K}uramoto to {C}rawford: exploring the onset of synchronization
  in populations of coupled oscillators.
\newblock {\em Phys. D: Non. Phe.}, 143:1--20, 2000.

\bibitem{zhai_kiss_05}
Y.M. Zhai, I.Z. Kiss, H.~Daido, and J.L. Hudson.
\newblock Extracting order parameters from global measurements with application
  to coupled electrochemical oscillators.
\newblock {\em Phys. D: Non. Phe.}, 205:57--69, 2005.

\bibitem{zhigulin_04}
P.Z. Zhigulin.
\newblock Dynamical motifs: building blocks of complex dynamics in sparsely
  connected random networks.
\newblock {\em Phy. Rev. Let.}, 92(23):238701, 2004.

\end{thebibliography}
